\documentclass[]{aa}  

\pdfoutput=1 

\usepackage{graphicx}
\usepackage{amsmath}
\usepackage{txfonts}
\usepackage{subfigure}
\usepackage{siunitx}
\usepackage{xspace}
\usepackage{natbib}
 
\usepackage{xcolor}

\bibpunct{(}{)}{;}{a}{}{,}

\newcommand*{\Euclid}{\textit{Euclid}\xspace}
\newcommand*{\diff}{\ensuremath{{\rm d}}}
\DeclareSIUnit \parsec{pc}
\DeclareSIUnit \h{\ensuremath{\mathnormal{h}}}
\DeclareSIUnit \hMpc{\per\h\mega\parsec}
\DeclareSIUnit \kmsMpc{\kilo\meter\per\second\per\mega\parsec}

\newcommand*{\degree}{\si{\degree}}
\newcommand*{\arcminute}{\si{\arcminute}}
\newcommand*{\hMpc}{\si{\hMpc}}
\newcommand*{\kmsMpc}{\si{\kmsMpc}}

\begin{document} 
\title{Going deep with Minkowski functionals of convergence maps}
\author{
Carolina Parroni\inst{1}
\and
Vincenzo F. Cardone\inst{1,4} 
\and
Roberto Maoli\inst{5}
\and
Roberto Scaramella\inst{1,4}}

\institute{
I.N.A.F. - Osservatorio Astronomico di Roma, via Frascati 33, 00040 - Monte Porzio Catone (Roma), Italy 
\and
Istituto Nazionale di Fisica Nucleare - Sezione di Roma 1, Piazzale Aldo Moro, 00185 - Roma, Italy
\and
Dipartimento di Fisica, Universit\`a di Roma "La Sapienza", Piazzale Aldo Moro, 00185 - Roma, Italy}

\abstract
{}
{Stage IV lensing surveys promise to make available an unprecedented amount of excellent data which will represent a huge leap in terms of both quantity and quality. This will open the way to the use of novel tools, which go beyond the standard second order statistics probing the high order properties of the convergence field. Motivated by these considerations, some of us (Vicinanza et al. 2019) have started a long term project aiming at using Minkowski Functionals (MFs) as complementary and supplementary probes to increase the lensing Figure of Merit (FoM).}
{As a second step along this path, we discuss the use of MFs for a survey made out of a wide total area $A_{\rm{tot}}$ imaged at a limiting magnitude $\rm{mag_{W}}$ containing a subset of area $A_{\rm{deep}}$ where observations are pushed to a deeper limiting magnitude $\rm{mag_{D}}$. We present an updated procedure to match the theoretically predicted MFs to the measured ones, taking into account the impact of map reconstruction from noisy shear data. We validate this renewed method against simulated data sets with different source redshift distributions and total number density, setting these quantities in accordance with the depth of the survey. We can then rely on a Fisher matrix analysis to forecast the improvement in the FoM due to the joint use of shear tomography and MFs under different assumptions on $(A_{\rm{tot}},\,A_{\rm{deep}},\,\rm{mag_{D}})$, and the prior on the MFs nuisance parameters.}
{It turns out that MFs can provide a valuable help in increasing the FoM of the lensing survey, provided the nuisance parameters are known with a non negligible precision. What is actually more interesting is the possibility to compensate for the loss of FoM due to a cut in the multipole range probed by shear tomography, which makes the results more robust against uncertainties in the modeling of nonlinearities. This makes MFs a promising tool to both increase the FoM and make the constraints on the cosmological parameters less affected by theoretical systematic effects.}
{}


\maketitle

\section{Introduction}

The concordance $\Lambda$CDM cosmological model, dominated by the dark energy driving cosmic speed up and the dark matter responsible for the clustering, assumes that the structures we observe today formed from gravitational instability and successive growth of the primordial fluctuations generated during the inflation epoch. Although on large scales the density field may be approximated as Gaussian, it is the non-Gaussianity on small scales to carry on additional information able to break some degeneracy among model parameters. 

Weak lensing (hereafter, WL) has proven to be an efficient tool to access such information. In any metric theory, light propagates along the geodesics of the metrics, which are determined by the matter distribution along the line of sight. As a consequence, lensing thus probes both the background expansion and the growth of structures, hence it is able to both put strong constraints on the dark energy equation of state and discriminate among general relativity and modified gravity. In the WL regime, lensing causes the distortion of the image of the emitting source, but this cosmic shear effect is so small that can only be detected statistically through the analysis of large sample of galaxies. Up to now, second order statistics have been used with the two-point correlation function and its Fourier counterparts, the power spectrum, has been considered  with remarkable results \citep[see, e.g.,][and refs. therein]{munshi2008, kilbinger2015, bartelmann2017}. 

The unprecedented amount of high quality data that Stage IV lensing surveys are expected to deliver will make it possible to deepen the analysis of the density field probing its non-Gaussianity. To this end, higher than second order statistics is needed. Among the different possibilities, Minkowski Functionals (hereafter, MFs) have already proven their reliability in the context of both CMB studies \citep{komatsu2003, eriksen2004, hikage2008, matsubara2010} and WL convergence maps \citep{matsubara2001, sato2001, taruya2002, matsubara2010, kratochvil2011, pratten2012, petri2013, shirasaki2014}. MFs are topological descriptors of the convergence field that depend on the complete set of higher order terms and multi-point correlation functions. In particular, non-Gaussianity manifests itself in deviations from the predictions for Gaussian random fields. Even cutting the perturbative expansion to the lowest order, such deviations are related to the convergence bispectrum (Fourier counterpart of the three-point correlation function) through the generalized skewness parameters. The need to go at large scales, however, asks for a detailed description of the nonlinearities affecting both the matter power spectrum and bispectrum. Moreover, in any practical application, the infinite series determining MFs deviations from the Gaussian case is truncated at the lowest order thus introducing a mismatch with the observed MFs even in the idealized case of noiseless maps. Needless to say, the presence of noise and the imperfect reconstruction of the convergence map from shear data make the theoretical predictions of MFs still more daunting. 

Motivated by these considerations, \citet{paper1} have first presented a calibration procedure to correct theoretical predictions for noiseless maps so that they match the MFs measured on reconstructed noisy convergence maps. We here propose a modification of their approach reducing the number of nuisance parameters starting from simplifying yet reasonable approximations. This offers us the possibility to improve the constraints on the cosmological parameters so that we repeat their Fisher matrix forecast analysis. We also investigate which survey strategy (e.g., wide and shallow or deep and narrow) is better suited to optimize the scientific return of MFs. Next generations surveys, both ground based as LSST  \citep{lsst2009} or on satellites as ESA \textit{Euclid} \citep{laureijs2011} and NASA WFIRST \citep{green2012}, will nevertheless rely on second order statistics so that it is mandatory that the optimization is carried out considering the combination of both cosmic shear tomography and MFs rather than single probes alone. We therefore consider several realistic combinations of area coverage and survey depth (keeping fixed the survey observation time) to generate simulated lognormal convergence fields, which are taken as input for the estimate of a reliable MFs data set with the corresponding covariance matrix.

  The paper is organized as follows: in Section \ref{simulations}, we describe how we obtained the set of simulated convergence maps, starting from the catalog generation for different survey depths, to the map reconstruction method used. In Section \ref{mf}, we introduce MFs and we describe how we performed the measurements on the simulated convergence maps, showing the results for some cases of interest. In Section \ref{mf_theo}, we describe MFs from a theoretical point of view, showing their connection to cosmology and presenting our new calibration procedure. In Section \ref{fisher}, we discuss the results we obtained from our Fisher matrix analysis, in terms of the Figure of Merit (FoM) improvement and survey optimization. In Section \ref{conclusions}, we draw our conclusions.  

\section{Convergence maps simulation} \label{simulations}

In \citet{paper1}, some of us have developed a calibration procedure to match the theoretical predictions for MFs measured on noiseless convergence fields with those estimated on reconstructed maps from noisy shear data. This has been validated using MICEv2.0 simulations, which cover a limited redshift range and model galaxies up to a limiting magnitude ($\rm{mag_{lim}} = 24.5$) shallower than the ones we are interested in here. We therefore need to produce a different set of simulated convergence maps to probe the extended redshift range which one is probing when going deeper in magnitude. Moreover, we want to mimic as close as possible what is expected for the \textit{Euclid} satellite mission, which means we need to input the same source redshift distribution. To this end, we use \texttt{FLASK} and the setting we describe in the following two subsections. 

\subsection{\texttt{FLASK} simulations}
	  
\texttt{FLASK} \citep[full-sky lognormal astro-fields simulation kit;][]{flask} is a public code designed to create two- or three-dimensional random realizations of different astrophysical fields, including weak lensing convergence and shear, reproducing the expected cross-correlations between the input fields. Such realizations follow a multivariate lognormal distribution, which, compared to a multivariate Gaussian distribution, results in a better approximation to the density and convergence fields, avoiding for example non-physical negative density values. Also, since for this study we are interested in capturing the non-Gaussian features contained in the convergence field, the lognormal distribution represents the simpler approximation that can convey this information. 

Its computational speed and flexibility makes \texttt{FLASK} strongly preferable with respect to full ray tracing or full N-body simulations. This is a key aspect since we need to simulate a large field area (to be split in a high number of patches) varying the source redshift distribution according to the limiting magnitude. We briefly outline the \texttt{FLASK} inner workings referring the reader to \citet{flask} for further details. 
	  
The code takes as input angular auto and cross power spectra calculated at a number of redshift slices provided by the user. These are then transformed to the real space to compute the associated Gaussian correlation functions to be transformed back to the harmonic space. Choleski decomposition is then used to generate Gaussian multipoles, which are the input for the creation of a \texttt{HEALPix} map whose pixels are exponentiated to obtain the associated lognormal fields. The user can then sample these fields according to its desired angular and radial selection functions thus mimicking the specifics of its desired survey. A catalog is finally generated assigning to each pixel a random angular position sampled within the pixel boundaries, and a redshift within its redshift slice. 

A caveat is in order when simulating correlated density and convergence fields. If one models the density field as a lognormal one, \texttt{FLASK} computes the convergence through an approximated line of sight integration obtained as a weighted Riemann sum of the simulated density in redshift bins. As a consequence of the small number of bins in the sum at low redshift, the resulting convergence field is not exactly lognormal. However, the corresponding power spectra reproduce the theoretical ones within $3\%$ for $z > 0.5$, while the precision quickly degrades at lower $z$. We will therefore impose a conservative cut $z > 0.55$ to select the sources to be included in our analysis.

We use \texttt{CLASS} \citep{blas2011,dio2013} to compute the input power spectra for 25 top-hat equispaced redshift bins over the range $0.0 \le z \le 2.5$ for a flat $\Lambda$CDM model with fiducial cosmological parameters
\begin{displaymath}
(\Omega_{\rm{M}},\,\Omega_{\rm{b}},\,h,\,n_{\rm{s}},\,\sigma_8) = (0.32,\,0.05,\,0.67,\,0.96,\,0.83)
\end{displaymath}
being $\Omega_{\rm{M}}$ ($\Omega_{\rm{b}}$) the present day value of the total matter (baryons only) density parameter, $h$ the Hubble constant (in units of $100\,\kmsMpc$), $n_{\rm{s}}$ the slope of the primordial power spectrum, and $\sigma_8$ the variance of the linear power spectrum smoothed over a top-hat window with size $R = 8\,\hMpc$. Nonlinearities at large $k$ are corrected for using the Halofit recipe \citep{smith2003,takahashi2012}. We also modify some keywords setting in \texttt{FLASK} with respect to the default ones setting $\rm{LRANGE} = 1-6000$, $\rm{SHEAR\_LMAX} = 2000$, and $\rm{NSIDE} = 2048$. We finally use a custom defined angular selection function to split the full-sky simulation in more manageable set of subfields.
   
\subsection{Survey depth and number density}

\texttt{FLASK} allows the user to input its own radial selection function so that both the total source number density and their redshift distribution match those of a given survey. Since we are interested in a Euclid-like survey, we set
\begin{equation}
n(z) = \frac{3 \, n_{\rm{g}}}{2 \, z_0} \, \left ( \frac{z}{z_0} \right )^2 \,\exp{ \left [ -\left ( \frac{z}{z_0} \right )^{3/2} \right ]}
\label{eq: nz}
\end{equation}
with $n_{\rm{g}}$ the number of galaxies per $\rm{arcmin}^{2}$, and $z_{0}  = z_{\rm{m}}/\sqrt{2}$ with $z_{\rm{m}}$ the median redshift.  
\begin{figure}
\centering
\includegraphics[scale=.50]{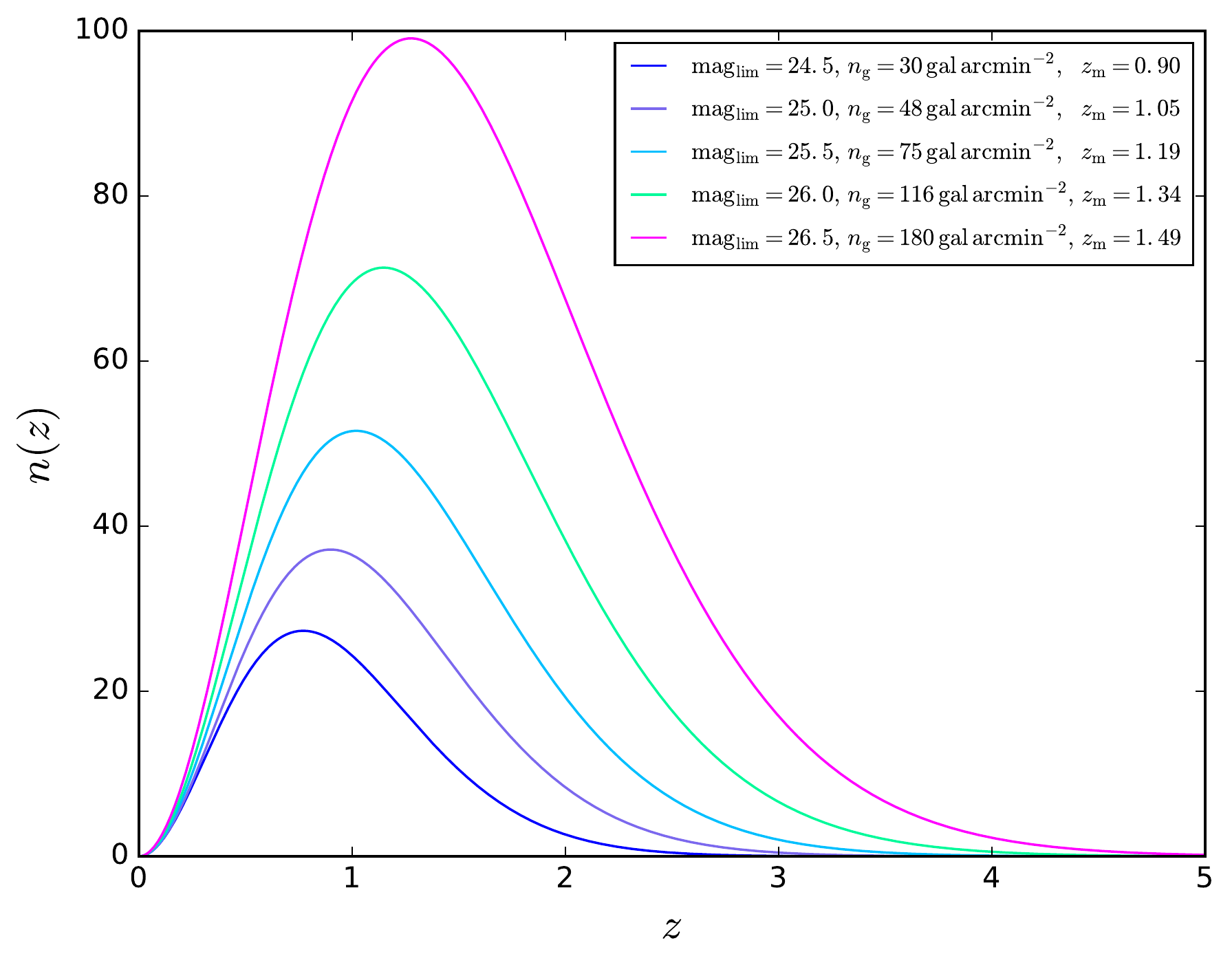}
\caption{Redshift selection functions used as input for \texttt{FLASK} to simulate different survey depths corresponding to different limiting magnitudes.}
\label{fig: nz}
\end{figure}
\begin{table}
\begin{center}
\begin{tabular}{ccc}
\hline \hline
$\rm{mag_{lim}}$ & $n_{\rm{g}}$ & $z_{\rm{m}}$ \\
\hline
24.5 & 30 & 0.90 \\
25.0 & 48 & 1.05 \\
25.5 & 75 & 1.19 \\
26.0 & 116 & 1.34 \\
26.5 & 180 & 1.49 \\
\hline
\end{tabular}
\end{center}
\caption{Limiting magnitude, total source number density (in ${\rm gal\,arcmin^{-2}}$), and median redshift.}
\label{tab: nz}
\end{table}
Both $n_{\rm{g}}$ (in ${\rm gal\,arcmin^{-2}}$) and $z_{\rm{m}}$ are function of the limiting magnitude $\rm{mag_{lim}}$, with $(n_{\rm{g}}, z_{\rm{m}}) = (30, 0.9)$ for $\rm{mag_{lim}} = 24.5$ for the wide area Euclid survey. We therefore need to model their scaling with $\rm{mag_{lim}}$, which we qualitatively do as follows. First, we note that \citet{hoekstra2017} has investigated the impact of undetected galaxies on the estimate of the shape measurement bias. To this end, they have modeled the dependence of the number of galaxies at a given $\rm{mag_{lim}}$ as a power law, which well approximates the number counts from the GEMS survey \citep{rix2004} in the F606W band and from the Hubble Ultra-Deep Field \citep[HUDF;][]{coe2006} in the F775W band. We integrated this power law up to the desired limiting magnitude thus getting the slope of the $n_{\rm{g}}$--$\rm{mag_{lim}}$ relation, while its amplitude is set so that it is $n_{\rm{g}}(\rm{mag_{lim}} = 24.5) = 30$ as for \Euclid. Note that such a rescaling is necessary since neither the F606W nor the F775W bands match the wide RIZ filter used by the Euclid imaging instrument. A similar rescaling is also used for the $z_{\rm{m}}$--$\rm{mag_{lim}}$ relation, whose behavior we obtain by interpolating the values in Table\,9 of \citet{coe2006} based on HUDF data. The values of $(n_{\rm{g}},\,z_{\rm{m}})$ thus obtained for the five different limiting magnitudes we consider are given in Table\,\ref{tab: nz}, while the corresponding redshift distributions $n(z)$ are shown in Fig.\,\ref{fig: nz}.	
\begin{figure}
\centering
\includegraphics[scale=.50]{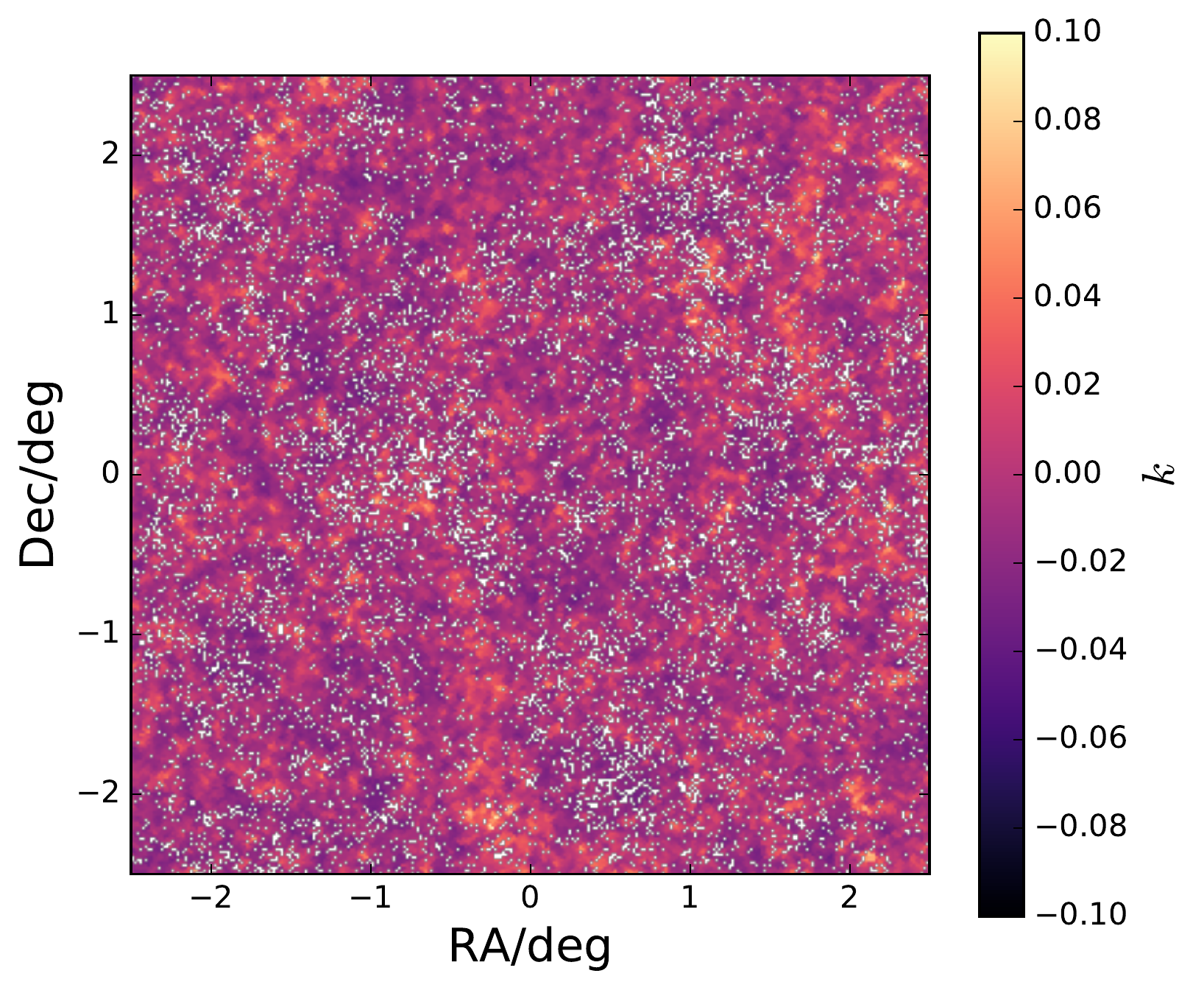} \\
\includegraphics[scale=.50]{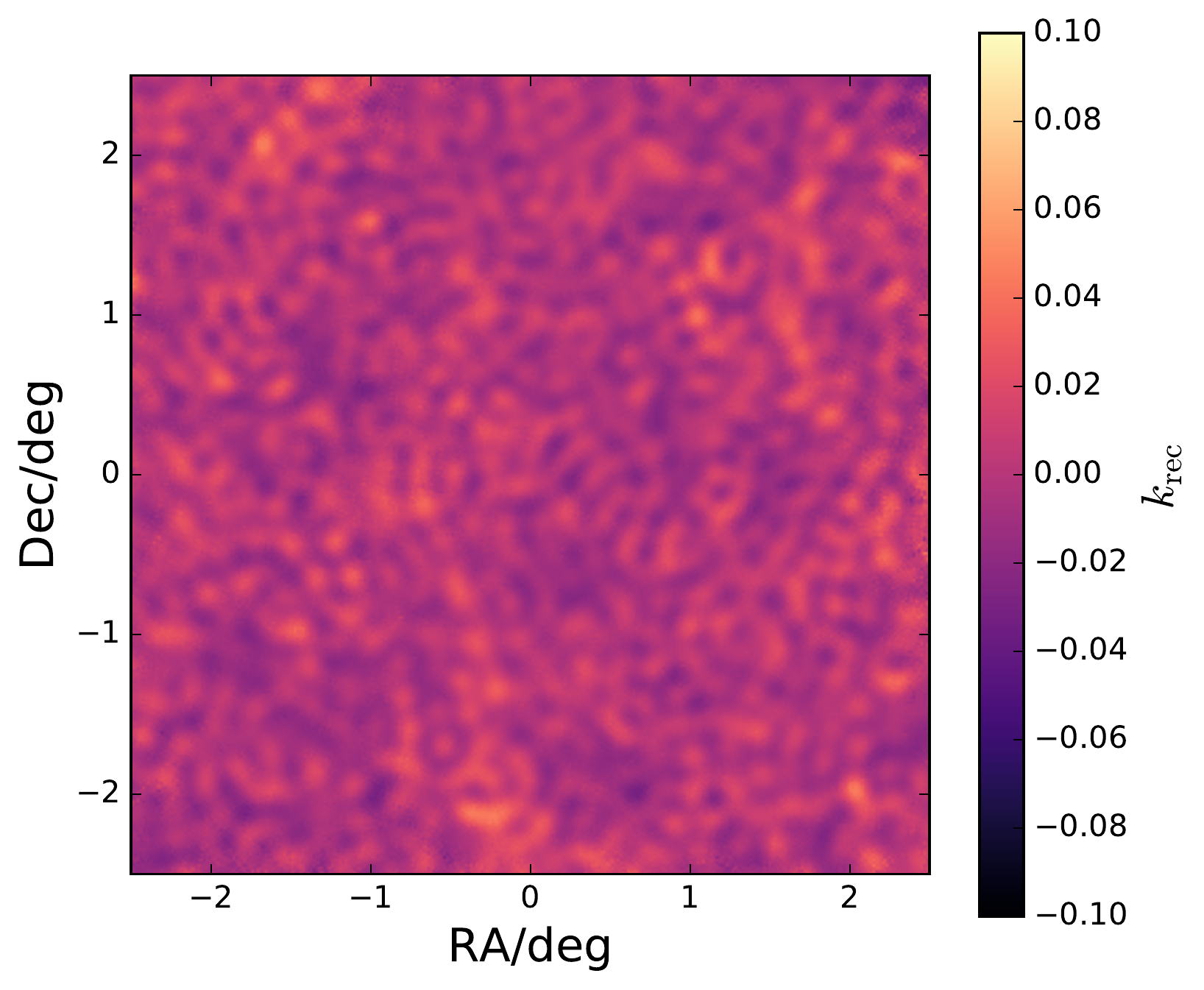}
\caption{\emph{Top}: convergence maps at $z = 0.9$ for a limiting magnitude $\rm{mag_{lim}} = 24.5$, simulated with \texttt{FLASK}. \emph{Bottom}: same map but KS reconstructed.}
\label{fig:k_rec}
\end{figure}
We remind the reader though that the calculation we made are based on the $i$ band magnitude, while \Euclid will provide imagining data in the broad RIZ band, which has currently never been used before for observations. Therefore, we do not expect our approximation to reproduce exactly \Euclid's redshift distribution but it will nevertheless allow us to illustrate the different results obtained changing the survey area and depth.

\subsection{Map reconstruction}

Running \texttt{FLASK} with the input parameters set as detailed above gives us catalogs with right ascension, declination, redshift $z$, convergence $\kappa$, and shear components $(\gamma_1,\,\gamma_2)$ for all the objects in the catalog. A gnomonic projection is then used to project on the plane of the sky under flat sky approximation, which holds for the $5 \times 5 \, \rm{deg}^2$ subfields we use. We also leave a gap of $\sim 1 \, \degree$ among two consecutive patches so that we can consider them as independent realization. The objects in each catalog are then split in redshift bins with equal width $\Delta z = 0.05$ and centered in $z$ from 0.5 to 1.8 in steps of 0.3 with the number density set according to the chosen limiting magnitude. We thus obtain \num{1108} independent convergence and shear maps for each redshift bin and limiting magnitude.

After smoothing the maps to $1\,\arcmin$ resolution, we add Gaussian noise to each pixel with variance fixed as \citep{hamana2004}:
\begin{equation}
\sigma^{2}_{\rm{pix}}=\frac{\sigma^{2}_{\rm{e}}}{2}\frac{1}{A_{\rm{pix}}\,n_{\rm{g}}}
\end{equation}  
with $\sigma_{\rm{e}}=0.3$ the intrinsic ellipticity, $A_{\rm{pix}}$ the pixel area, and $n_{\rm{g}}$ the number density of galaxies. Note that, since $n_{\rm{g}}$ increases with $\rm{mag_{lim}}$, deeper maps will be less noisy as expected. In order to mimic what is done in actual data analysis, we reconstruct the convergence maps from the noisy shear data using different methods. After comparing with simulated convergence maps, we finally opt for a variant of the popular KS method \citep{ks1993} modified to account for the impact of systematic effects such as projection effects and masking \citep{pires2009,jullo2014}. Fig. \ref{fig:k_rec} shows, as an example, a convergence map at $z = 0.9$ for $\rm{mag_{lim}} = 24.5$. On the top, the original map obtained with the simulated convergence and, on the bottom, the same map reconstructed with the method outlined above. Hereafter, whenever we will mention convergence maps, we will always refer to the set of reconstructed maps.

\section{Minkowki functionals: measurement} \label{mf}

Let us consider a smooth two-dimensional random field $k(x, \, y)$ with zero mean and variance $\sigma_0^2$. We first define the excursion set $Q_{\nu}$ as the region where the normalized field $k/\sigma_0$ is larger than a given threshold $\nu$. We can then define the three MFs as 
\begin{equation}
V_0(\nu) = \frac{1}{A} \, \int_{Q_{\nu}}{\diff a} \; ,
\label{eq: V0def}
\end{equation}
\begin{equation}
V_1(\nu) = \frac{1}{4 \, A} \int_{\partial Q_{\nu}}{\diff l} \; ,
\label{eq: V1def}
\end{equation}
\begin{equation}
V_2(\nu) = \frac{1}{2 \, \pi \, A} \int_{\partial Q_{\nu}}{{\diff l \; \cal{K}}} \ ,
\label{eq: V2def}
\end{equation}
where $A$ is the map area, $\partial Q_{\nu}$ the excursion set boundary, $\diff a$ and $\diff l$ the surface and line element along $\partial Q_{\nu}$, and ${\cal{K}}$ its curvature. $(V_0, \, V_1, \, V_2)$ are the area, the perimeter, and the genus characteristics (i.e., the number of connected regions above a given $\nu$ minus that of connected regions below $\nu$) of the excursion set $Q_{\nu}$. MFs can be redefined in a more convenient way as
\begin{equation}
V_{0}(\nu) = \frac{1}{A} \int_{A}{\diff x \diff y \; \Theta(\kappa- \nu \, \sigma_0)} \; ,
\label{eq: V0meas}
\end{equation}
\begin{equation}
V_{1}(\nu) = \frac{1}{4 \, A} \int_{A}{\diff x \diff y \; \delta_D(\kappa - \nu \, \sigma_0) \, \sqrt{\kappa_{x}^{2} + \kappa_{y}^{2}}} \ ,
\label{eq: V1meas}
\end{equation}
\begin{equation}
V_{2}(\nu) = \frac{1}{2 \, \pi \, A}
\int_{A}{\diff x \diff y \; \delta_D(\kappa - \nu \, \sigma_0) 
\frac{2 \, \kappa_x \, \kappa_y \, \kappa_{xy} - \kappa_{x}^{2} \, \kappa_{yy} - \kappa_{y}^{2} \, \kappa_{xx}}
{\kappa_{x}^{2} + \kappa_{y}^{2}}} \; ,
\label{eq: V2meas}
\end{equation}
where we have explicitly considered the case of the convergence field $\kappa(x, \, y)$ and expressed the threshold as a multiple of its variance $\sigma_0$. In Eqs.(\ref{eq: V0meas})--(\ref{eq: V2meas}), it is $\kappa_i = \partial \kappa/\partial x_i$, and $\kappa_{ij} = \partial^2 \kappa/\partial x_i \partial x_j$ with $(i, \, j) = (x, \, y)$, i.e., MFs are computed in terms of the field and its derivatives. Using these definitions, it is then straightforward to implement an algorithm to measure the MFs from the map. One should, however, deal with numerical issues coming from the conversion of integrals into discrete sums, derivatives into finite differences, and the Dirac-$\delta$ into discrete $\nu$ binning.
\begin{figure}
\centering
\includegraphics[scale=.50]{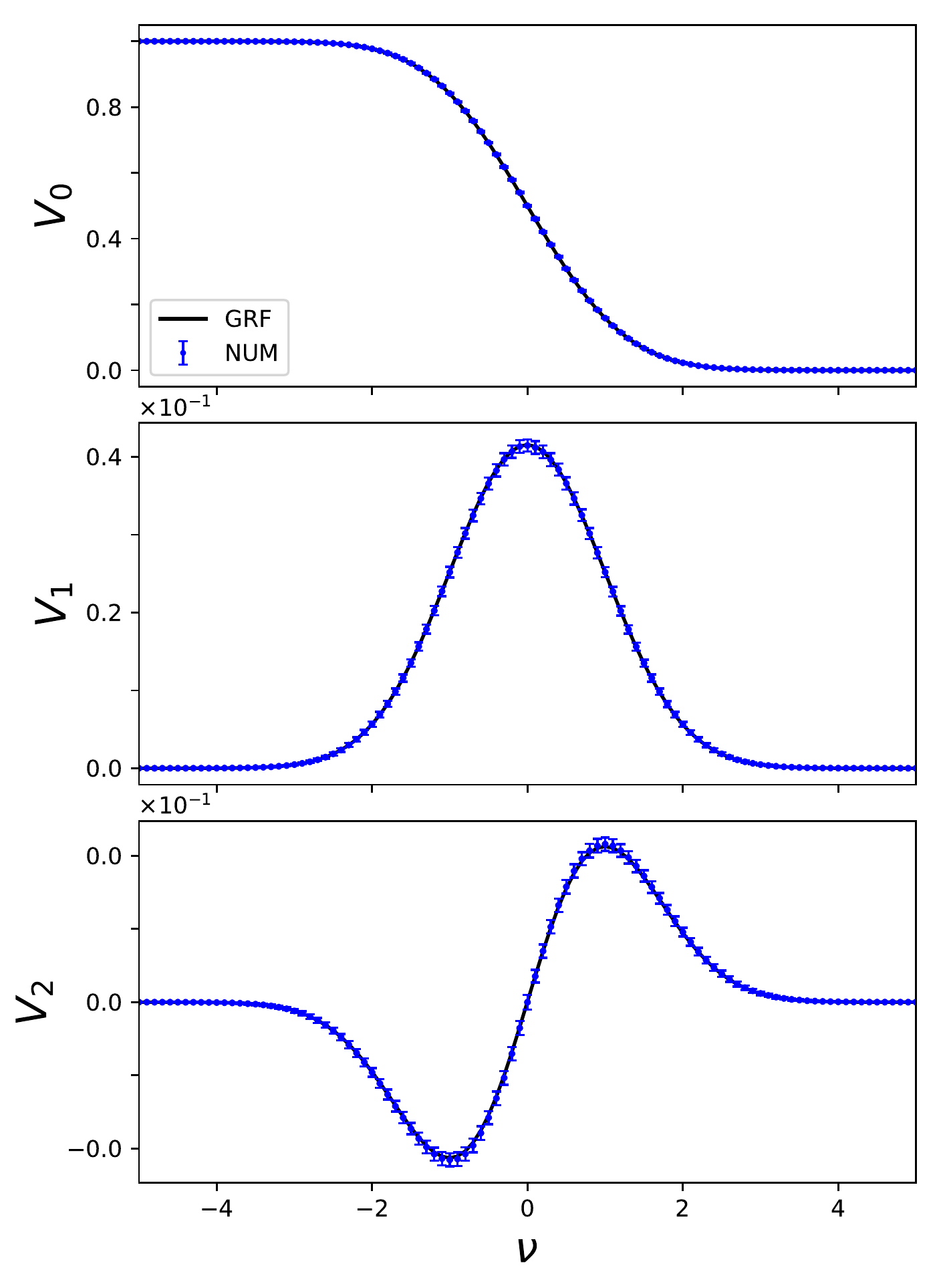}
\caption{\small Numerical (blue dots) vs expected (solid black line) MFs from 500 realizations of a Gaussian random field.}
\label{fig:grf_mfs}
\end{figure}
In order to validate our pipeline, we realized 500 random Gaussian maps that we input to our code for measuring the MFs. We then take the mean as final estimate, and the standard deviation as uncertainty, finally getting the results in Fig.\,\ref{fig:grf_mfs} where the solid black line is the theoretical prediction (see later). The measured $(V_0, \, V_1)$ deviate from the theoretical expectation at $\nu=2$, the threshold use for the rest of the analysis, less than $1\%$, while the discrepancy is slightly larger (up to $1\%$) for $V_2$. This is related to the way the theoretical value is computed since it relies on the values of $(\sigma_0, \, \sigma_1)$, defined later, which are themselves measured on the maps. We therefore do not ascribe this larger difference to a missing ingredient in the theoretical estimate thus deeming as reliable our measurement pipeline for all MFs. 

We then measure MFs on the \num{1108} simulated convergence maps varying the redshift bin centers $z$, the limiting magnitude $\rm{mag_{lim}}$, and the scale $\theta_{\rm{s}}$ of the Gaussian filter used to smooth the maps before MFs estimate. In particular, we consider
\begin{displaymath}
\left \{ 
\begin{array}{ll}
\displaystyle{0.6 \le z \le 1.8} & \displaystyle{{\rm in \ steps \ of \ } \Delta z = 0.3} \\
 & \\
\displaystyle{24.5 \le \rm{mag_{lim}} \le 26.5} & \displaystyle{{\rm in \ steps \ of \ } \Delta \rm{mag_{lim}} = 0.5} \\
 & \\
\displaystyle{\ang{;2.0;} \le \theta_{\rm{s}} \le \ang{;14.0;}} & \displaystyle{{\rm in \ steps \ of \ } \Delta \theta_{\rm{s}} = \ang{;4.0;} } \\
\end{array}
\right . 
\end{displaymath}
\begin{figure*}
\centering
\includegraphics[scale=.50]{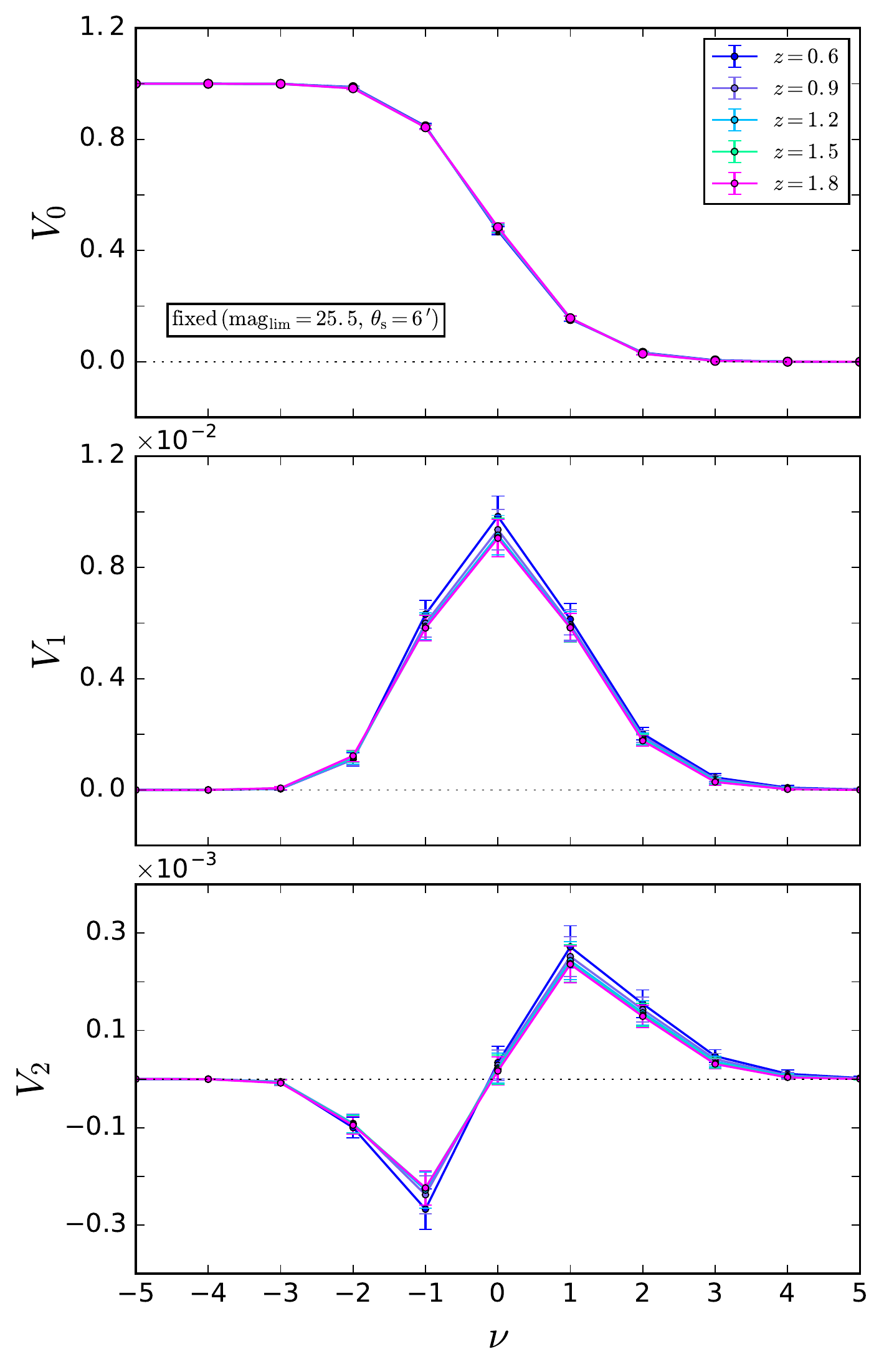}
\includegraphics[scale=.50]{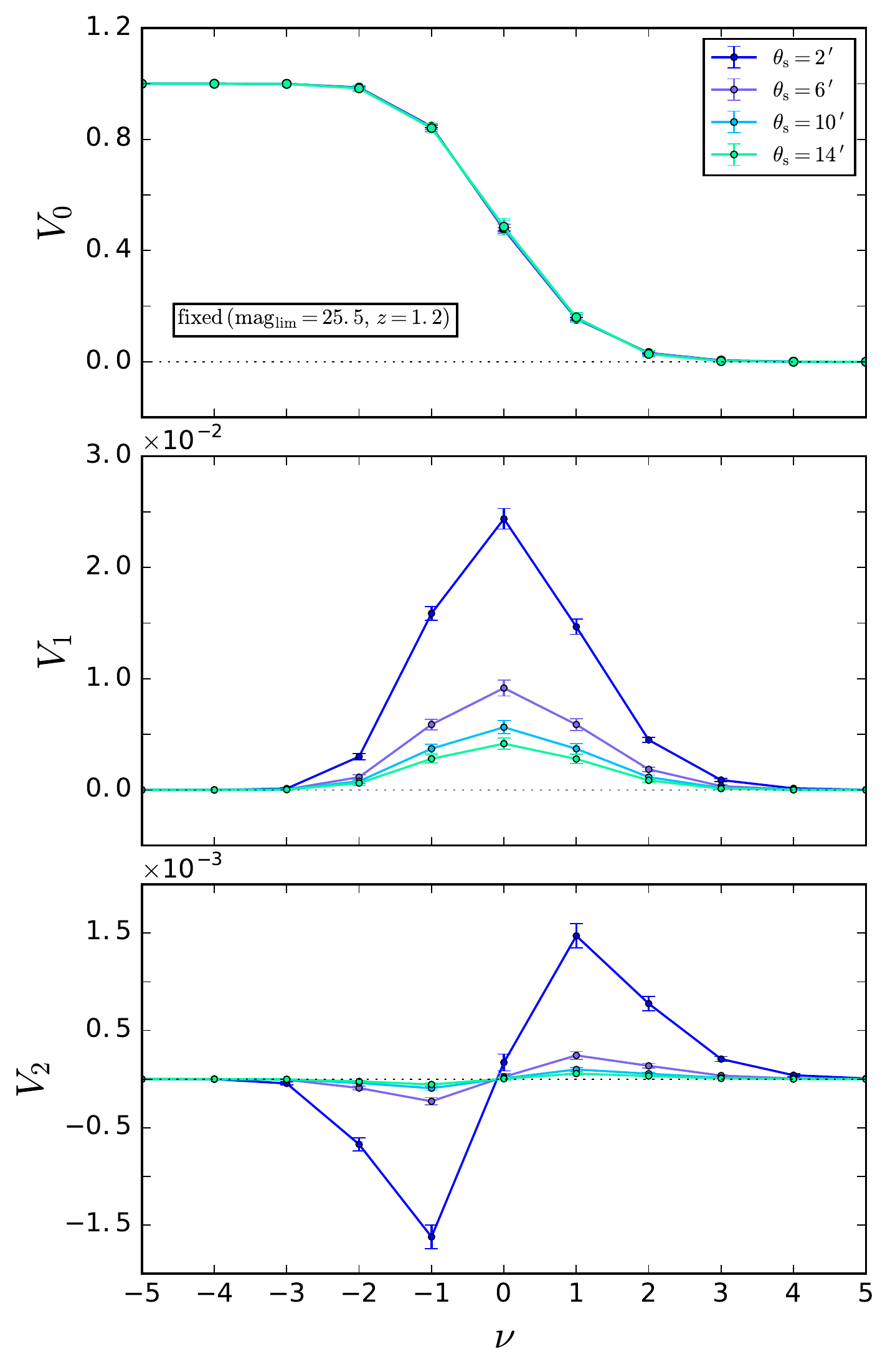}
 \caption{\emph{Left}: MFs as measured from maps with $\rm{mag_{lim}} = 25.5$ as function of the threshold $\nu$ for different values of the redshift bin center $z$ and fixed smoothing scale ($\theta_{\rm{s}} = \ang{;6;}$). \emph{Right}: same as in the left panel but for different values of the smoothing scale $\theta_{\rm{s}}$ and fixed redshift bin center ($z = 1.2$).}
\label{fig:25.5}
\end{figure*}
\begin{figure}
\centering
\includegraphics[scale=.50]{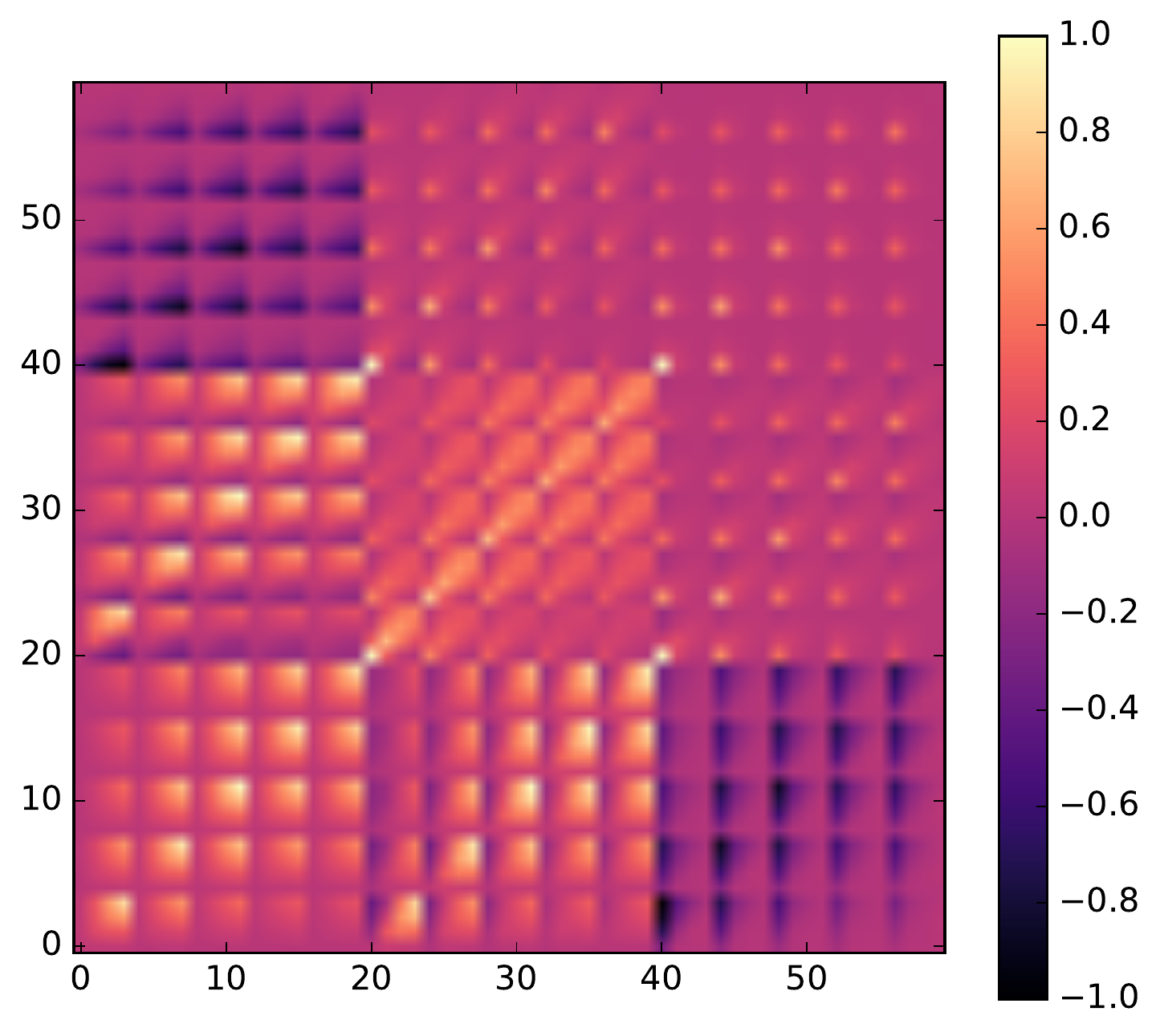}
 \caption{Normalized covariance matrix for the MFs data vector ${\bf D}$ defined in the text for the case with $\rm{mag_{lim}} = 25.5$.}
\label{fig:mfs_cov}
\end{figure}
It is instructive to look at Fig.\,\ref{fig:25.5}, which shows the three MFs as function of the S/N ratio $\nu$ for the illustrative case of a survey with $\rm{mag_{lim}} = 25.5$ (other cases being qualitatively similar). For fixed smoothing scale (left panels), the overall scaling with $\nu$ is the same, with the redshift value only entering to determine the MF amplitude. In particular, differences in $V_0$ are typically quite small being no larger than $\sim 2\%$, while they increase up to $\sim 8\%$ ($\sim 20\%$) for $V_1$ ($V_2$). A similar argument holds for the dependence on the smoothing angle for fixed $\theta_{\rm{s}}$, with differences that can now be easily appreciated as shown by the results in the right panel. Such results suggest that it is not the behavior of MFs with $\nu$ that carries the relevant information but rather the dependence on the redshift and the smoothing angle. We will therefore set $\nu = 2$ in the rest of the analysis, referring the reader to the next section for the reason of this particular value.

Our observed MFs data vector will then be: 
\begin{equation}
\mathbf{D}_{\rm{obs}} = \mathbf{D}_{0}+\mathbf{D}_{1}+\mathbf{D}_{2}
\end{equation}
with
\begin{align}
\mathbf{D}_{n}  = & \{ V_{n}(0.6, \, \ang{;2;}), V_{n}(0.6, \, \ang{;6;}), V_{n}(0.6, \, \ang{;10;}), V_{n}(0.6, \, \ang{;14;}) \} \nonumber \\
\cup & \{ V_{n}(0.9, \, \ang{;2;}), V_{n}(0.9, \, \ang{;6;}), V_{n}(0.9, \, \ang{;10;}), V_{n}(0.9, \, \ang{;14;}) \} \nonumber \\
\cup & \{ V_{n}(1.2, \, \ang{;2;}), V_{n}(1.2, \, \ang{;6;}), V_{n}(1.2, \, \ang{;10;}), V_{n}(1.2, \, \ang{;14;}) \} \nonumber \\
 \cup & \{ V_{n}(1.5, \, \ang{;2;}), V_{n}(1.5, \, \ang{;6;}), V_{n}(1.5, \, \ang{;10;}), V_{n}(1.5, \, \ang{;14;}) \} \nonumber \\
\cup & \{ V_{n}(1.8, \, \ang{;2;}), V_{n}(1.8, \, \ang{;6;}), V_{n}(1.8, \, \ang{;10;}), V_{n}(1.8, \, \ang{;14;}) \} \nonumber 
 \end{align}
where the values are computed as the mean over the \num{1108} convergence maps realized for each given $\rm{mag_{lim}}$. The covariance matrix can then be estimated as  
\begin{equation}
\mathbf{Cov}_{ij}^{\rm{obs}} = \frac{\sum_{k=1}^{N_{\rm{maps}}} \, {\left[\mathbf{D}_{\rm{obs}}^{i, \, k}-\mathbf{D}_{\rm{obs}}^{i}\right] \, \left[\mathbf{D}_{\rm{obs}}^{j, \, k}-\mathbf{D}_{\rm{obs}}^{j}\right]}}{N_{\rm{maps}}-1}
\label{eq: covmatdata}
\end{equation}
where $N_{\rm{maps}} = \num{1108}$ is the total number of convergence maps, $\mathbf{D}_{\rm{obs}}^{i, \, k}$ is the $i$th component of the data vector, calculated on the $k$th map, and $\mathbf{D}_{\rm{obs}}^{i}$ is the same component averaged over all maps. In Fig. \ref{fig:mfs_cov} we show the normalized covariance matrix obtained in the case $\rm{mag_{lim}} = 25.5$, as an example. We notice that for $V_{0}$ the correlation increases with the smoothing scale, while it seems quite insensitive to the redshift. On the other hand, for $V_{1}$ and $V_{2}$, we see higher correlations for small values of $\theta_{\rm{s}}$ and $z$. We observe the same pattern for the cross correlation between $V_{1}$ and $V_{2}$ and, while $V_{0}$ and $V_{1}$ appear correlated for larger $\theta_{\rm{s}}$, $V_{0}$ and $V_{2}$ result to be anticorrelated. The strong correlations that we find among some elements of the data vector suggests that one can actually reduce the dimension of ${\bf D}$ without losing appreciable information. We will therefore investigate this possibility too. 
\begin{figure*}
\centering
\includegraphics[scale=.55]{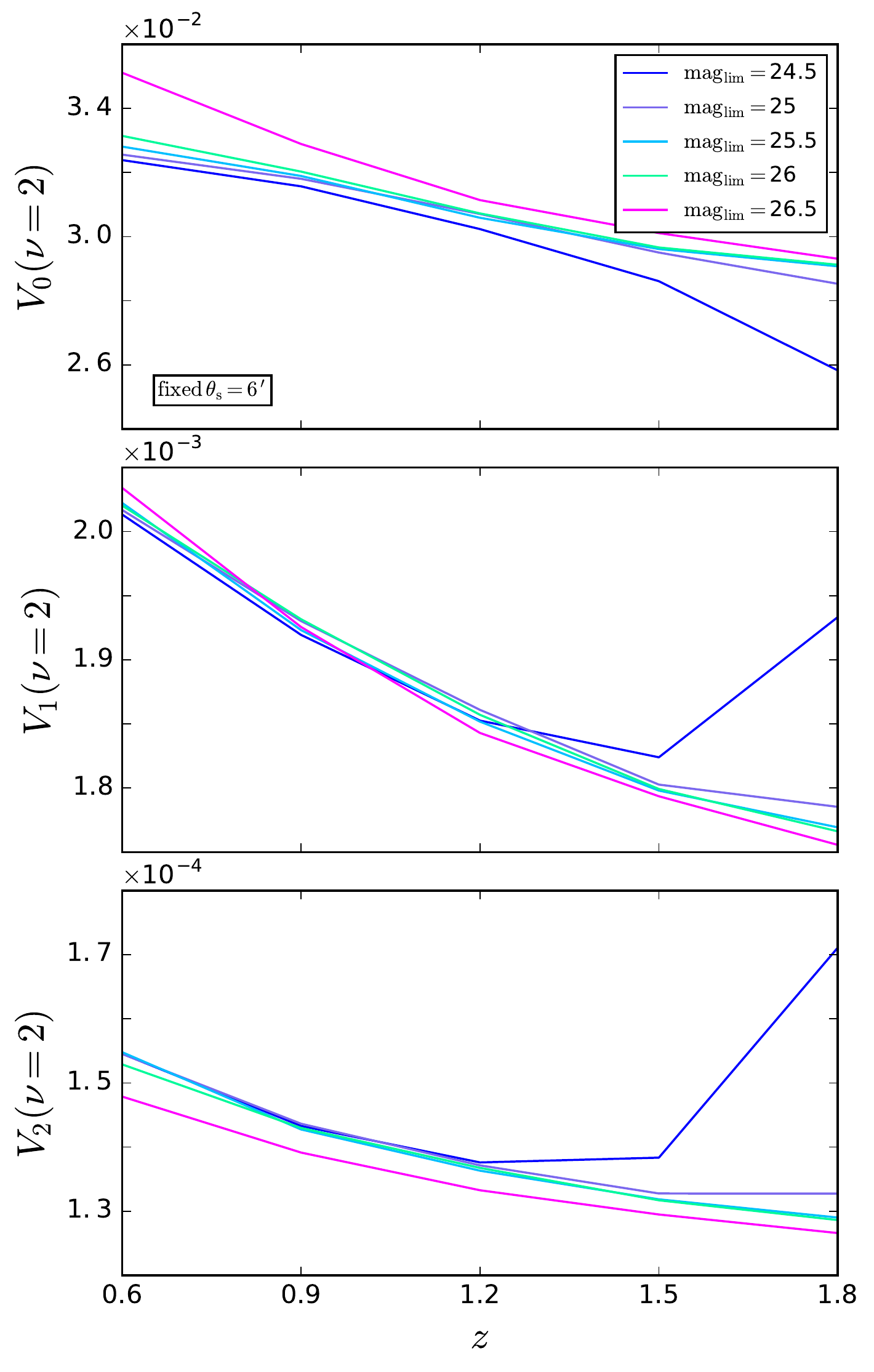}
\includegraphics[scale=.55]{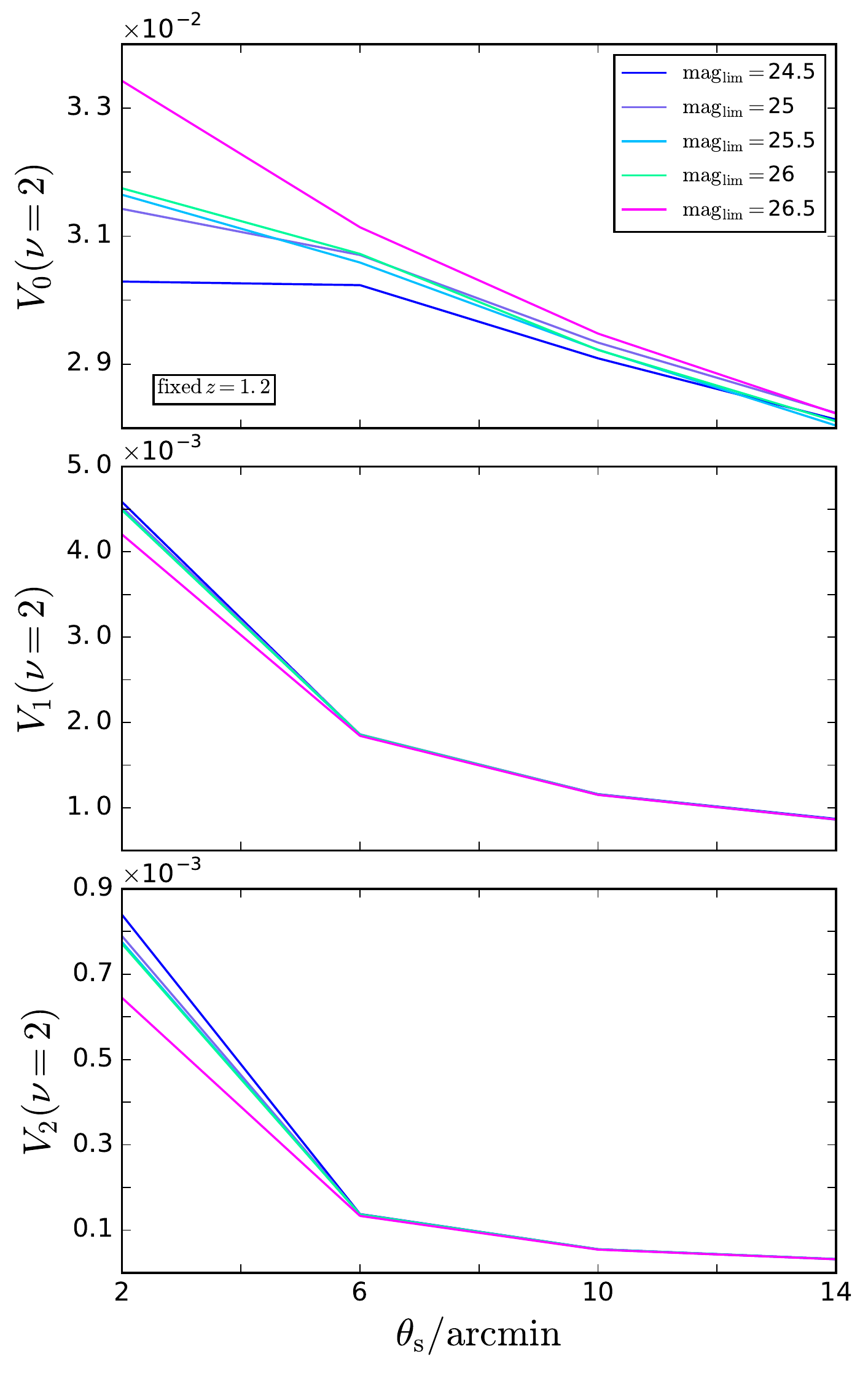}
\caption{\emph{Left}: MFs for $\nu = 2$ and different values of the limiting magnitude $\rm{mag_{lim}}$ as a function of the redshift bin center $z$, with fixed smoothing scale ($\theta_{\rm{s}} = \ang{;6;}$). \emph{Right}: same as in the left panel but as a function of the smoothing angle $\theta_{\rm{s}}$ with fixed redshift bin center ($z = 1.2$).}
\label{fig:mfs_nu2}
\end{figure*}
Let us now focus on the case $\nu = 2$ to investigate how the MFs change as function of the limiting magnitude. This is shown in Fig.\,\ref{fig:mfs_nu2} where we plot MFs as function of $z$ and $\theta_{\rm{s}}$ for five different $\rm{mag_{lim}}$. For a fixed smoothing angle, the difference among the MFs amplitude at different $\rm{mag_{lim}}$ tends to decrease with $z$ being no larger than $\sim 9\%$. The only remarkable exception is the case with $\rm{mag_{lim}} = 24.5$, which gives a $\sim 35\%$ difference in the $V_2$ amplitude at large $z$. However, this is actually a consequence of the small number of galaxies in the high $z$ bins for a survey as shallow as the one with this limiting magnitude. As a consequence, the map reconstruction becomes more noisy and less reliable, and should be corrected for as we demonstrate in the next section. Right panels show that the dependence on $\rm{mag_{lim}}$ is less and less important as the smoothing angle $\theta_{\rm{s}}$ increases. This is expected since the larger $\theta_{\rm{s}}$, the more the convergence field is Gaussian so that the MFs depend on $\nu$ only. As a consequence, we get the unfortunate result that the scales leading the most of the information are the ones at small $\theta_{\rm{s}}$, which are at the same time the noisiest ones. Next section discusses how to mitigate the impact of noise through a suitable calibration procedure.
      
\section{Minkowski functionals: theoretical predictions} \label{mf_theo}

In order for an observable to be of any use in constraining cosmological parameters, it is mandatory to have a way to theoretically compute its expected value. This is analytically possible for MFs only in the case of Gaussian random fields, while deviation from non-Gaussianity (as the ones for the convergence field) can be dealt with in an approximated way through a perturbative series expansion. Such a method, however, does not take into account systematic effects introduced by imperfect map reconstruction from noisy shear data. In \citet{paper1}, some of us have shown how to successfully account for this through a semianalytical approach calibrated on simulations. We will below first summarize the main steps and results, and then present a simplified yet still reliable way to reduce the number of nuisance parameters. 

\subsection{Minkowski fucntionals for noiseless convergence fields}

For a Gaussain random field, MFs can be exactly computed as \citep{adler1981,tomita1986}
\begin{eqnarray}
V_{n}^{\rm G}(\nu)  & = & \frac{1}{(2 \, \pi)^{(n + 1)/2}} \, \frac{\omega_2}{\omega_{2 - n} \, \omega_n} \, 
\left ( \frac{\sigma_1}{\sqrt{2} \, \sigma_0} \right )^{n} \nonumber \\
 & \times & \exp{\left ( - \frac{\nu^2}{2} \right )} \, {\cal{H}}_{n-1}(\nu)
\label{eq: GaussMFs}
\end{eqnarray}
with $\omega_n=\pi^{n/2} \, \left[\Gamma(n/2+1)\right]^{-1}$ so that it is $\omega_0 = 1$, $\omega_1 = 2$, $\omega_2 = \pi$. Here, it is assumed that the field has null mean,  variance $\sigma_0$, and variance of its covariant derivative $\sigma_1$, while ${\rm{\cal{H}}}_n(\nu)$ are Hermite polynomials.

If the field is only mildly non-Gaussian, a perturbative expansion can be used hence writing
\begin{equation}
V_{n}(\nu) = V_{n}^{\rm G}(\nu) + \delta V_{n}(\nu)\; .
\label{eq: mfsum}
\end{equation}
The deviation from the Gaussian prediction can be expanded in terms of $\sigma_0 = \langle \kappa^2 \rangle$ as 
\begin{eqnarray}
\delta V_{n}(\nu) & = & 
\frac{1}{(2 \, \pi)^{(n + 1)/2}} \, \frac{\omega_2}{\omega_{2 - n} \, \omega_n} \, 
\left ( \frac{\sigma_1}{\sqrt{2} \, \sigma_0} \right )^{n} \,\exp{\left ( - \frac{\nu^2}{2} \right )} \nonumber \\
 & \times & \left [ \delta V_n^{(2)}(\nu) \, \sigma_0 + \delta V_{n}^{(3)}(\nu) \, \sigma_0^2 + \ldots \right ]
\label{eq: deltavn}
\end{eqnarray}
with $\sigma_{1}^{2} = \left \langle (\nabla \kappa)^2 \right \rangle$. To the lowest order in $\sigma_0$, the coefficient of the correction term reads
\begin{eqnarray}
\delta V_{n}^{(2)}(\nu) & = &
S^{(0)} \, {\cal{H}}_{n + 2}(\nu)/6 \nonumber \\
 & + &  n\, S^{(1)}\, {\cal{H}}_{n}(\nu)/3  \nonumber \\
 & + & n\, (n - 1)\, S^{(2)}\, {\cal{H}}_{n - 2}(\nu)/6 \; ,
\label{eq: deltaV2}
\end{eqnarray}
where $S^{(n)}$ are generalized skewness quantities defined from of the convergence field and its derivatives
\begin{eqnarray}
S^{(0)} & = & \frac{\langle \kappa^3 \rangle}{\sigma_{0}^{4}} \; , \\
S^{(1)} & = & -\frac{3}{4}\, \frac{\langle \kappa^2 \, \nabla^2 \kappa \rangle}{\sigma_{0}^{2} \, \sigma_{1}^{2}} \; , \\
S^{(2)} & = & - 3 \, \frac{\langle (\nabla \kappa) \cdot (\nabla \kappa) \, ( \nabla^2 \kappa) \rangle}{\sigma_{1}^{4}} \; .
\label{eq: skewdef}
\end{eqnarray}
Both the variance terms $\sigma_n$ and the the generalized skewness parameters $S^{(n)}$ can be expressed in terms of the polyspectra of the field. For the variances, it is indeed \citet{munshi2012}
\begin{equation}
\sigma_n^2 = \frac{1}{4 \, \pi}\, \sum_{\ell}{(2 \, \ell + 1) \, [\ell \, (\ell + 1)]^n \, {\cal{C}}(\ell) \, {\cal{W}}^{2}(\ell)}
\label{eq: variance}
\end{equation}
where ${\cal{C}}(\ell)$ is the lensing convergence power spectrum for sources at redshift $z_{\rm s}$, and ${\cal{W}}(\ell)$ is the Fourier transform of the smoothing filter. The cosmological information is contained in ${\cal{C}}(\ell)$, which is given by
\begin{equation}
{\cal{C}}(\ell) = \frac{c}{H_0} \int_{0}^{z_{\rm s}}
{\diff z \; \frac{W^2(z)}{r^2(z) \, E(z)} \, P_{\rm NL}\left [ \frac{\ell}{\chi(z)}, \, z \right ]}
\label{eq: defconvps}
\end{equation}
with 
\begin{equation}
W(z) = \frac{3}{2} \, \Omega_{\rm M} \, \left ( \frac{H_0}{c} \right )^2 \, \chi(z) \, \left [1 - \frac{\chi(z)}{\chi(z_{\rm s})} \right ] \; ,
\label{eq: defkernel}
\end{equation}
where $E(z) = H(z)/H_0$ is the dimensionless Hubble function, $\chi(z)$ the comoving distance, $r(z)$ the comoving angular diameter distance, and $P_{\rm NL}(k, \, z)$ the nonlinear matter power spectrum evaluated in $k = \ell/\chi(z)$ because of the Limber approximation. Note that hereafter we will assume a spatially flat universe. We will use a Gaussian filter to smooth the map, i.e., 
\begin{equation}
{\cal{W}}(\ell) = \exp{\left ( - \ell^2 \, \sigma_{\rm s}^2 \right )}
\label{eq: wlgauss}
\end{equation}
with $\sigma_{\rm s}$ the smoothing length. Generalized skewness quantities (that are connected with third-order moments) can be expressed as
\begin{equation}
S^{(n)} = \sum_{\ell}{(2 \, \ell + 1) \, {\cal{S}}^{(n)}(\ell)}
\label{eq: sncalc}
\end{equation}
where, adopting a compact notation, we get
\begin{equation}
{\cal{S}}^{(n)}(\ell) =  
\sum_{\ell_1, \ell_2}{
\frac{s_n \, (\ell, \, \ell_1, \, \ell_2) \, {\cal{B}}(\ell, \, \ell_1, \, \ell_2) \, \tilde{{\cal{W}}}(\ell, \, \ell_1, \, \ell_2)}
{K_{n}(\sigma_0, \, \sigma_1)}} \; ,
\label{eq: snelle}
\end{equation}
with 
\begin{equation}
K_{n}(\sigma_0, \, \sigma_1) = \left \{ 
\begin{array}{ll}
12 \, \pi \, \sigma_{0}^{4} & \ \ n = 0 \\ 
 & \\
16 \, \pi \, \sigma_{0}^{2} \, \sigma_{1}^{2} & \ \ n = 1 \\
 & \\
8 \, \pi \, \sigma_{1}^{4} & \ \ n = 2 \\
\end{array}
\right . \ ,
\label{eq: normcst}
\end{equation}
and 
\begin{eqnarray}
s_0 & = & 1  \nonumber \\
 & \nonumber \\ 
s_1 & = & \ell \, (\ell + 1) + \ell_1 \, (\ell_1 + 1) + \ell_2 \, (\ell_2 + 1) \nonumber \\
 & \nonumber \\
s_2 & = & [\ell \, (\ell + 1) + \ell_1 \, (\ell_1 + 1) - \ell_2 \, (\ell_2 + 1)] \, \ell_2\, (\ell_2 + 1) + {\rm cp} \nonumber 
\end{eqnarray}
where cp denotes cyclic permutation. In Eq.(\ref{eq: snelle}), the cosmological information is coded into the convergence bispectrum 
\begin{eqnarray}
{\cal{B}}(\ell_1, \, \ell_2, \, \ell_3) & = & \frac{c}{H_0}  \\ 
 & \times & \int_{0}^{z_{\rm{s}}}
{\diff z \; \frac{W^3(z)}{r^4(z) \, E(z)} \, B_{\rm NL}\left [ \frac{\ell_1}{\chi(z)}, \, \frac{\ell_2}{\chi(z)}, \, \frac{\ell_3}{\chi(z)} \right ]} \nonumber 
\label{eq: convbps}
\end{eqnarray}
with $B_{\rm NL}(k_1, \, k_2, \, k_3, \, z)$, the matter bispectrum, evaluated at $k_i = \ell_i/\chi(z)$ because of the Limber approximation. The contribution of each multipole to the sum in Eq.(\ref{eq: snelle}) is weighted by 
\begin{equation}
{\tilde{\cal{W}}}(\ell_1, \, \ell_2, \, \ell_3) = {\cal{J}}(\ell_1, \, \ell_2, \, \ell_3) \, {\cal{W}}(\ell_1) \, {\cal{W}}(\ell_2) \, {\cal{W}}(\ell_3)
\label{eq: defwtilde}
\end{equation}
with
\begin{equation}
{\cal{J}}(\ell_1, \, \ell_2, \, \ell_3) = \frac{{\cal{I}}^2(\ell_1, \, \ell_2, \, \ell_3)}{2 \, \ell_3 + 1} \; ,
\label{eq: defjelle}
\end{equation}
and
\begin{equation}
{\cal{I}} = \sqrt{\frac{(2 \, \ell_1 + 1) \, (2 \, \ell_2 + 1) \, (2 \, \ell_3 + 1)}{4 \, \pi}}
\left (
\begin{array}{lll}
\ell_1 & \ell_2 & \ell_3 \\
 & & \\
0 & 0 & 0 \\
\end{array}
\right ) \; .
\label{eq: defielle}
\end{equation}
Note that the presence of the Wigner-3j symbols accounts for the fact that only triangular configurations (i.e., $\vec{k}_1 + \vec{k}_2 + \vec{k}_3 = 0$) contribute to the sum. 

\subsection{Observable Minkowski fucntionals}

As yet hinted above, Eqs.(\ref{eq: mfsum})--(\ref{eq: deltaV2}) refer to the case of a  noiseless convergence field. However, there are different reasons why they cannot be straightforwardly used to predict the MFs, which are measured on actual convergence maps. First, $\kappa$ is not a directly observed quantity, but it is rather reconstructed from the shear data so that multiplicative and additive bias are present. Second, the field is also shifted from its true value because of the noise. In \citet{paper1}, we have addressed this issue postulating that at the lowest order these effects can be described as 
\begin{equation}
\kappa_{\rm{obs}} = (1 + m_{\kappa}) \, \kappa + N
\label{eq: kappalin}
\end{equation}
with $m_{\kappa}$ the multiplicative bias, and $N$ the zero mean noise. Starting from Eq.(\ref{eq: kappalin}) and assuming the noise is not correlated with the signal, it is possible to propagate the effect of noise and bias on the variance of the field and its derivatives, and on the generalized skewness parameters. One finally ends up with the following expressions for the observable MFs
\begin{eqnarray}
V_{0, \, {\rm obs}}(\nu) & = & \frac{1}{\sqrt{2 \, \pi}}\, \exp{\left ( - \frac{\nu^2}{2} \right )} \nonumber \\
 & \times & \Bigg \{ {\cal{H}}_{-1}(\nu) + 
\left [ (1 + m_{\kappa})^2 + {\cal{R}}_{0}^{2} \right ]^{1/2} \, \sigma_0  \nonumber \\
 & \times &  \frac{[(1 + m_{\kappa})^3 \, S^{(0)} + \tilde{S}^{(0)}] \, {\cal{H}}_2(\nu)}
{6 \, [(1 + m_{\kappa})^2 + {\cal{R}}_{0}^{2}]^2} \Bigg \} \; , 
\label{eq: v0obs}
\end{eqnarray}
\begin{eqnarray}
V_{1, \, {\rm obs}}(\nu) & = & \frac{1}{8}\, \left ( \frac{\sigma_1}{\sqrt{2} \, \sigma_0} \right ) \, \exp{\left ( - \frac{\nu^2}{2} \right )} \nonumber \\
 & \times & \left [ \frac{(1 + m_{\kappa})^2 + {\cal{R}}_{1}^{2}}{(1 + m_{\kappa})^2 + {\cal{R}}_{0}^{2}} 
\right ]^{1/2} \nonumber \\
 & \times & \Bigg \{ {\cal{H}}_{0}(\nu) + \left [ (1 + m_{\kappa})^2 + {\cal{R}}_{0}^{2} \right ]^{1/2} \, \sigma_0  \nonumber \\
 & \times & \left [ \frac{[(1 + m_{\kappa})^3 \, S^{(0)} + \tilde{S}^{(0)}] \, {\cal{H}}_3(\nu)}
{6 \, [(1 + m_{\kappa})^2 + {\cal{R}}_{0}^{2}]^2} \right . \nonumber \\
 & + & \left . \frac{[(1 + m_{\kappa})^3 \, S^{(1)} + \tilde{S}^{(1)}] \, {\cal{H}}_{1}(\nu)}
{3 \, [(1 + m_{\kappa})^2 + {\cal{R}}_{0}^{2}] \, [(1 + m_{\kappa})^2 + {\cal{R}}_{1}^{2}]} \right ] \Bigg \} \ , 
\label{eq: v1obs}
\end{eqnarray}
\begin{eqnarray}
V_{2, \, {\rm obs}}(\nu) & = & \frac{1}{(2 \, \pi)^{3/2}}\, \left ( \frac{\sigma_1}{\sqrt{2} \, \sigma_0} \right )^2 \, \exp{\left ( - \frac{\nu^2}{2} \right )} \nonumber \\
 & \times & \left [ \frac{(1 + m_{\kappa})^2 + {\cal{R}}_{1}^{2}}{(1 + m_{\kappa})^2 + {\cal{R}}_{0}^{2}} 
\right ] \nonumber \\
 & \times & \Bigg \{ {\cal{H}}_{1}(\nu) + \left [ (1 + m_{\kappa})^2 + {\cal{R}}_{0}^{2} \right ]^{1/2} \, \sigma_0  \nonumber \\
 & \times & \left [ \frac{[(1 + m_{\kappa})^3 \, S^{(0)} + \tilde{S}^{(0)}] \, {\cal{H}}_4(\nu)}
{6 \, [(1 + m_{\kappa})^2 + {\cal{R}}_{0}^{2}]^2} \right . \nonumber \\
 & + & \frac{2 \, [(1 + m_{\kappa})^3 \, S^{(1)} + \tilde{S}^{(1)}] \, {\cal{H}}_{2}(\nu)}
{3 \, [(1 + m_{\kappa})^2 + {\cal{R}}_{0}^{2}] \, [(1 + m_{\kappa})^2 + {\cal{R}}_{1}^{2}]}  \nonumber \\
 & + & \left . \frac{[(1 + m_{\kappa})^3 \, S^{(2)} + \tilde{S}^{(2)}] \, {\cal{H}}_0(\nu)}
{3 \, [(1 + m_{\kappa})^2 + {\cal{R}}_{1}^{2}]^2} \right ] \Bigg \} \; , 
 \label{eq: v2obs}
\end{eqnarray}
where, to shorten the notation, we have introduced the variance ratios ${\cal{R}}_i = \sigma_{i\rm{N}}/\sigma_i$, and defined the tilde skewness parameters as
\begin{equation}
\tilde{S}^{(0)} = S^{(0)}_{\rm{N}} \, {\cal{R}}_{0}^{4} \; ,
\label{eq: s0tilde}
\end{equation}
\begin{equation}
\tilde{S}^{(1)} = S^{(1)}_{\rm{N}} \, {\cal{R}}_{0}^{2} \, {\cal{R}}_{1}^{2} 
- (3/4) \, \sigma_{21}^{2} \, \left [(1 + m_{\kappa}) \, {\cal{R}}_{2}^{2} + {\cal{R}}_{0}^{2} \right ] \; ,
\label{eq: s1tilde}
\end{equation}
\begin{equation}
\tilde{S}^{(2)} = S^{(2)}_{\rm{N}} \, {\cal{R}}_{1}^{4} 
- 3 \, (1 + m_{\kappa}) \, \sigma_{21}^{2} \, \left [(1 + m_{\kappa}) \, {\cal{R}}_{2}^{2}  + {\cal{R}}_{1}^{2} \right ] \; ,
\label{eq: s2tilde} 
\end{equation}
with $\sigma_{21} = \sigma_{2}/\sigma_{1}$, and the label N denoting noise related quantities. Eqs.(\ref{eq: v0obs})--(\ref{eq: s2tilde}) make it possible to estimate the MFs of the observed convergence field in terms of the variances $\sigma_{n}$ and generalized skewness parameters $S^{(n)}$ (with $n = 0, 1, 2$) of both the actual convergence field and the noise. 

\subsection{Validation and calibration}

Eqs.(\ref{eq: v0obs})--(\ref{eq: v2obs}) have been obtained under some assumptions, which, although reasonable, are nevertheless only approximations. It is therefore mandatory to validate them by fitting to measured MFs in the simulated data set. Such a test will also give us the fiducial values of the nuisance parameters entering them so that we refer to the full process as `calibration'. \citet{paper1} have successfully calibrated these relations against the MICEv2 catalog data. To this end, they modeled the functions ${\cal{R}}_n$ and $\tilde{S}^{n}$ as power laws of the redshift and smoothing scales thus summing up to a total of 13 nuisance parameters ${\bf p}_{\rm{nuis}}$. 

We have reconsidered this problem here carefully looking at how these quantities are defined thus finding out a way to reduce the dimensionality of ${\bf p}_{\rm{nuis}}$ without significantly degrading the overall fit quality. To this end, let us first look at the variance ratios given by
\begin{equation}
 {\cal{R}}_n = \frac{\sigma_{n\rm{N}}(\theta_{\rm{s}})}{\sigma_{n}(\theta_{\rm{s}})} = 
{\cal{R}}_{n}^{\rm{ref}} \left [ \frac{\sigma_{n\rm{N}}(\theta_{\rm{s}})}{\sigma_{n\rm{N}}(\theta_{\rm{ref}})} \right ] \, \left [ \frac{\sigma_{n}(\theta_{\rm{s}}, \, z)}{\sigma_{n\rm{N}}(\theta_{\rm{ref}}, \, z_{\rm{ref}})} \right ]^{-1} \; ,
\label{eq: varratio}
\end{equation}
where the label ref denotes a quantity evaluated at some arbitrary chosen reference values of $(\theta_{\rm{s}}, \, z)$, which we fix as $(\ang{;2;}, \, 0.3)$. Based on their own definition, both terms in square parentheses have a predictable scaling once the cosmological model is given. As a consequence, we can reduce the number of nuisance parameters \footnote{This is different from \citet{paper1} where we instead modeled the dependence on $\theta_{\rm{s}}$ as a power law thus adding three more parameters to fix the slopes.} to only three, namely the values $({\cal{R}}_{0}^{\rm{ref}}, \, {\cal{R}}_{1}^{\rm{ref}}, \, {\cal{R}}_{2}^{\rm{ref}})$ of the ratios at the reference point. 

Let us now consider the $\tilde{S}^{(n)}$ quantities starting from the case $n = 0$, which we can conveniently rearrange as follows
\begin{eqnarray}
\tilde{S}^{(0)}(\theta_{\rm{s}}, \, z) & = & S^{(0)}_{\rm{N}}(\theta_{\rm{s}}) \, {\cal{R}}_{0}^{4} \nonumber \\
 & = &  S^{(0)}(\theta_{\rm{s}}, \,  z) \, \left [ \frac{S^{(0)}_{\rm{N}}(\theta_{\rm{s}})}{S^{(0)}(\theta_{\rm{s}}, \,  z)} \right ] \, \left [ \frac{\sigma_{0\rm{N}}(\theta_{\rm{s}})}{\sigma_{0}(\theta_{\rm{s}}, \,  z)} \right ]^4 \nonumber \\
 & = &  S^{(0)}(\theta_{\rm{s}}, \,  z_{\rm{ref}})  
\left [ \frac{\sigma_{0}(\theta_{\rm{s}}, \,  z_{\rm{ref}})}{\sigma_{0}(\theta_{\rm{s}}, \,  z)} \right ]^4 \, \beta_{0}(\theta_{\rm{s}}, \,  z_{\rm{ref}})
\label{eq: szskew} 
\end{eqnarray}
where we have used Eqs.(\ref{eq: sncalc})--(\ref{eq: normcst}), and defined
\begin{equation}
\beta_{0} = \frac{\sum_{\ell}{(2 \, \ell + 1) \, \sum_{\ell_1, \, \ell_2}{{\cal{T}}_{\rm{N}}(\ell, \, \ell_1, \, \ell_2)}}}
{\sum_{\ell}{(2 \, \ell + 1) \, \sum_{\ell_1, \, \ell_2}{{\cal{T}}(\ell, \, \ell_1, \, \ell_2)}}} \nonumber
\label{eq: defepszero}
\end{equation}
with
\begin{equation}
{\cal{T}}_{\rm{N}}(\ell, \, \ell_1, \, \ell_2) = s_0(\ell, \, \ell_1, \, \ell_2) \, {\cal{B}}_{\rm{N}}(\ell, \, \ell_1, \, \ell_2) \, {\cal{W}}(\ell, \, \ell_1, \, \ell_2) \; ,
\end{equation}

\begin{equation}
{\cal{T}}(\ell, \, \ell_1, \, \ell_2) = s_0(\ell, \, \ell_1, \, \ell_2) \, {\cal{B}}(\ell, \, \ell_1, \, \ell_2, \, z_{\rm{ref}}) \, {\cal{W}}(\ell, \, \ell_1, \, \ell_2) \; ,
\end{equation}
where ${\cal{B}}_{\rm{N}}(\ell, \, \ell_1, \, \ell_2)$ is the noise bispectrum. It is worth noting that both the numerator and denominator depend on the smoothing scale $\theta_{\rm{s}}$ only through the same weight function ${\cal{W}}(\ell, \, \ell_1, \, \ell_2)$ so that we can argue that their ratio is weakly dependent on it. As a working assumption, we will therefore consider $\beta_{0}$ independent on $\theta_{\rm{s}}$ and redefine it as\footnote{We scale $\beta_{0}$ with respect to ${\cal{R}}_{0}^{\rm{ref}}$ just to have a reference dimensionless value, but this choice is actually arbitrary. As a consequence, there is no reason to expect $\beta_{0}^{\rm{ref}}$ to be a small number.} 
\begin{displaymath}
\beta_{0} = \beta_{0}^{\rm{ref}} \, {\cal{R}}_{0}^{\rm{ref}}
\end{displaymath}
so that Eq.(\ref{eq: szskew}) finally reads
\begin{equation}
\tilde{S}^{(0)}_{\rm{N}}(\theta_{\rm{s}}, \, z) = \beta_{0}^{\rm{ref}} \, {\cal{R}}_{0}^{\rm{ref}} \, S^{(0)}(\theta_{\rm{s}}, \, z_{\rm{ref}})
\, \left [ \frac{\sigma_{0}(\theta_{\rm{s}}, \, z_{\rm{ref}})}{\sigma_{0}(\theta_{\rm{s}}, \, z)} \right ]^4   \ .
\label{eq: skewzeroend}
\end{equation}
Proceeding in a similar way, we also get
\begin{eqnarray}
\tilde{S}^{(1)}_{\rm{N}} & = & \beta_{1}^{\rm{ref}} \, {\cal{R}}_{1}^{\rm{ref}} \, S^{(1)}(\theta_{\rm{s}}, \, z_{\rm{ref}})
\, \left [ \frac{\sigma_{0}(\theta_{\rm{s}}, \, z_{\rm{ref}})}{\sigma_{0}(\theta_{\rm{s}}, \, z)} \right ]^2   
\, \left [ \frac{\sigma_{1}(\theta_{\rm{s}}, \, z_{\rm{ref}})}{\sigma_{1}(\theta_{\rm{s}}, \, z)} \right ]^2   \nonumber \\
 & - & \frac{3}{4} \, \sigma_{21}^{2}(\theta_{\rm{s}}, \, z) 
\, \left [(1 + m_{\kappa}) \, {\cal{R}}_{2}^{2}(\theta_{\rm{s}}, \, z) + {\cal{R}}_{0}^{2}(\theta_{\rm{s}}, \, z) \right ]
\label{eq: skewunoend}
\end{eqnarray}
\begin{eqnarray}
\tilde{S}^{(2)}_{\rm{N}} & = & \beta_{2}^{\rm{ref}} \, {\cal{R}}_{2}^{\rm{ref}} \, S^{(2)}(\theta_{\rm{s}}, \, z_{\rm{ref}})
\, \left [ \frac{\sigma_{1}(\theta_{\rm{s}}, \, z_{\rm{ref}})}{\sigma_{1}(\theta_{\rm{s}}, \, z)} \right ]^4   \\
 & - & 3 \, (1 + m_{\kappa}) \, \sigma_{21}^{2}(\theta_{\rm{s}}, \, z) 
\, \left [(1 + m_{\kappa}) \, {\cal{R}}_{2}^{2}(\theta_{\rm{s}}, \, z) + {\cal{R}}_{1}^{2}(\theta_{\rm{s}}, \, z) \right ] \nonumber
\label{eq: skewdueend}
\end{eqnarray}
so that now all the quantities entering $\tilde{S}^{(n)}$ have a predictable dependence on $(\theta_{\rm{s}}, \, z)$, and one is left with three additional nuisance parameters, namely $\left (\beta_{0}^{\rm{ref}}, \, \beta_{1}^{\rm{ref}}, \, \beta_{2}^{\rm{ref}} \right )$.

We finally end up with the following nuisance parameters:
\begin{displaymath}
{\bf p}_{\rm{nuis}} = \left \{m_{\kappa}, \,
{\cal{R}}_{0}^{\rm{ref}}, \, {\cal{R}}_{1}^{\rm{ref}}, \, {\cal{R}}_{2}^{\rm{ref}}, \, 
\beta_{0}^{\rm{ref}}, \, \beta_{1}^{\rm{ref}}, \, \beta_{2}^{\rm{ref}} \right \}
\end{displaymath}
which is definitely smaller than in \citet{paper1}, being ${\bf p}_{\rm{nuis}}$ a 7 rather than 13-dimensional vector. This reduction has been made possible by having fixed the way the variance ratios ${\cal{R}}_i$ and the the noise skewness related quantities $\tilde{S}^{(n)}$ scale with $(\theta_{\rm{s}}, \, z)$. We therefore need to validate these calibration approach by fitting to the mock data set we have constructed from the simulated convergence maps at different depths. 

We do this by a straightforward fitting procedure, i.e., we minimize a pseudo\,-\,$\chi^2$ merit function defined as

\begin{equation}
\chi^2({\bf p}_{\rm{nuis}}) = ({\bf D}_{\rm{obs}} - {\bf D}_{\rm{th}}) ({\bf Cov}^{\rm{obs}})^{-1} ({\bf D}_{\rm{obs}} - {\bf D}_{\rm{th}})^T
\label{eq: chisqcal}
\end{equation}
with ${\bf D}_{\rm{obs}}$ and ${\bf D}_{\rm{th}}$ the observed and theoretically predicted MF dataset, and ${\bf Cov}^{\rm{obs}}$ the corresponding covariance matrix as determined from the simulated maps.

It is worth stressing that the vector of nuisance parameters ${\bf p}_{\rm{nuis}}$ changes according to which dataset is considered. For instance, as shown by Eq.(\ref{eq: v0obs}), should one include $V_0$ only, ${\bf p}_{\rm{nuis}}$ would reduce to $ \left \{m_{\kappa}, \, {\cal{R}}_{0}^{\rm{\rm{ref}}}, \, \beta_{0}^{\rm{\rm{ref}}} \right \}$. We must therefore repeat the fit for each MFs combination with the consequence that the same nuisance parameter can have different fiducial values depending on which dataset one is considering. Similarly, the errors on the parameters will be different, which will impact the estimate of the systematics covariance matrix discussed later.

A cautionary remark is in order here. Compared to Paper I, we have changed the calibration method in two major aspects. First, we make a single joint fit to the full dataset, rather than first fitting to $V_0$ only and then to $(V_1, V_2)$. Second, we now include the full covariance matrix ${\bf Cov}^{\rm{obs}}$ in the $\chi^2$ function, hence taking care of the correlation among the components of the dataset. In Paper I, we only considered the diagonal elements since that procedure allowed us to better minimize the scatter of the residuals of single MFs. This is no more the case with the revised calibration formulae we have introduced here so that we prefer to adopt the present more statistically correct approach.

We perform the calibration for the different mock data set varying the limiting magnitude and the dataset used (i.e., whether we include only one MF or a combination of them). Compared to \citet{paper1}, the performance of the calibration is similar, although we note a small increase of $\rho_{0}^{\rm{RMS}}$ when $V_0$ it is not used in combination with other MFs. A straightforward comparison is, however, not possible because of the radically different fitting procedure, the larger redshift range (up to 1.8 instead of 1.4) and a different source redshift distribution. We also note that a marked decrease in $\rho_{n}^{\rm{RMS}}$ could be obtained cutting the lowest redshift bin, which is at the edge of the redshift range recommended by \texttt{FLASK} authors. Cutting MFs at $z = 0.6$ would reduce the number of observables thus decreasing the overall constraining power of this probe. Future lensing surveys will, on the contrary, will not be affected by this problem so that also MFs at such low $z$ will be usable. That is why we have preferred to retain these terms in the data vector at the cost of an increase of the RMS of best fit residuals with respect to what will likely be available when using fully realistic mock data for calibration. We therefore expect our results to err on the side of conservativeness. 

The MCMC method we use to explore the nuisance parameter space allows us to sample the joint posterior that we then use to propagate the errors on ${\bf p}_{\rm{nuis}}$ on the MFs. We thus obtained a covariance matrix, which represents the uncertainties we would have on the MFs even if they had been measured with infinite precision.  In a sense, this is the uncertainty coming from our imperfect theoretical modeling of the MFs and the lack of knowledge of the exact nuisance parameters. Put in other words, this is what we refer to as the systematics covariance matrix which we will denote as ${\bf Cov}^{\rm{sys}}$. We stress that ${\bf Cov}^{\rm{sys}}$ will depend on the fitting so that it is different for each MFs dataset.

\section{Fisher matrix forecasts} \label{fisher}

Eqs.(\ref{eq: v0obs})--(\ref{eq: v2obs}) allow us to match theoretical and measured MFs correcting for the overall impact of missing higher order terms, imperfect reconstruction from the shear field, and noise in the ellipticity data. As input, one needs to specify the cosmological model parameters 
\begin{displaymath}
{\bf p}_{\rm{cosmo}} = \left \{ \Omega_{\rm{M}}, \, \Omega_{\rm{b}}, \, w_0, w_a, \, h, \, n_{\rm{s}}, \, \sigma_8 \right \}
\end{displaymath}
and the nuisance ones
\begin{displaymath}
{\bf p}_{\rm{nuis}} = \left \{m_{\kappa}, \,
{\cal{R}}_{0}^{\rm{ref}}, \, {\cal{R}}_{1}^{\rm{ref}}, \, {\cal{R}}_{2}^{\rm{ref}}, \, 
\log{\beta_{0}^{\rm{ref}}}, \, \log{\beta_{1}^{\rm{ref}}}, \, \log{\beta_{2}^{\rm{ref}}} \right \}
\end{displaymath}
where we have changed to logarithmic units for $\beta_{n}^{\rm{ref}}$ since this quantity may change over a order of magnitude wide range.
Fitting simulated data sets mimicking as close as possible the actual data can help constraining ${\bf p}_{\rm{nuis}}$, but it is a safer option to left them free to vary to account for possible missing ingredients in the simulations. As a consequence, we do not expect MFs alone to be able to constrain the full set of parameters so that, in the following, we will always consider the joint use of MFs and the standard cosmic shear tomography using the Fisher matrix formalism \citep{tegmark1997} to make forecasts. 

This analysis has been already presented in \citet{paper1} for a survey mimicking the redshift distribution of the MICEv2 catalog and using a larger number of nuisance parameters. We address here a complementary issue. Planned future surveys will typically cover a wide area to a relatively shallow limiting magnitude, and a narrow region to a deeper limiting magnitude. We therefore investigate how the survey performances improve when one uses the shear tomography and MFs measured on the wide area and MFs from the deeper region. Assuming independence of the probes, the total Fisher matrix will read
\begin{figure*}
\centering
\includegraphics[scale=.40]{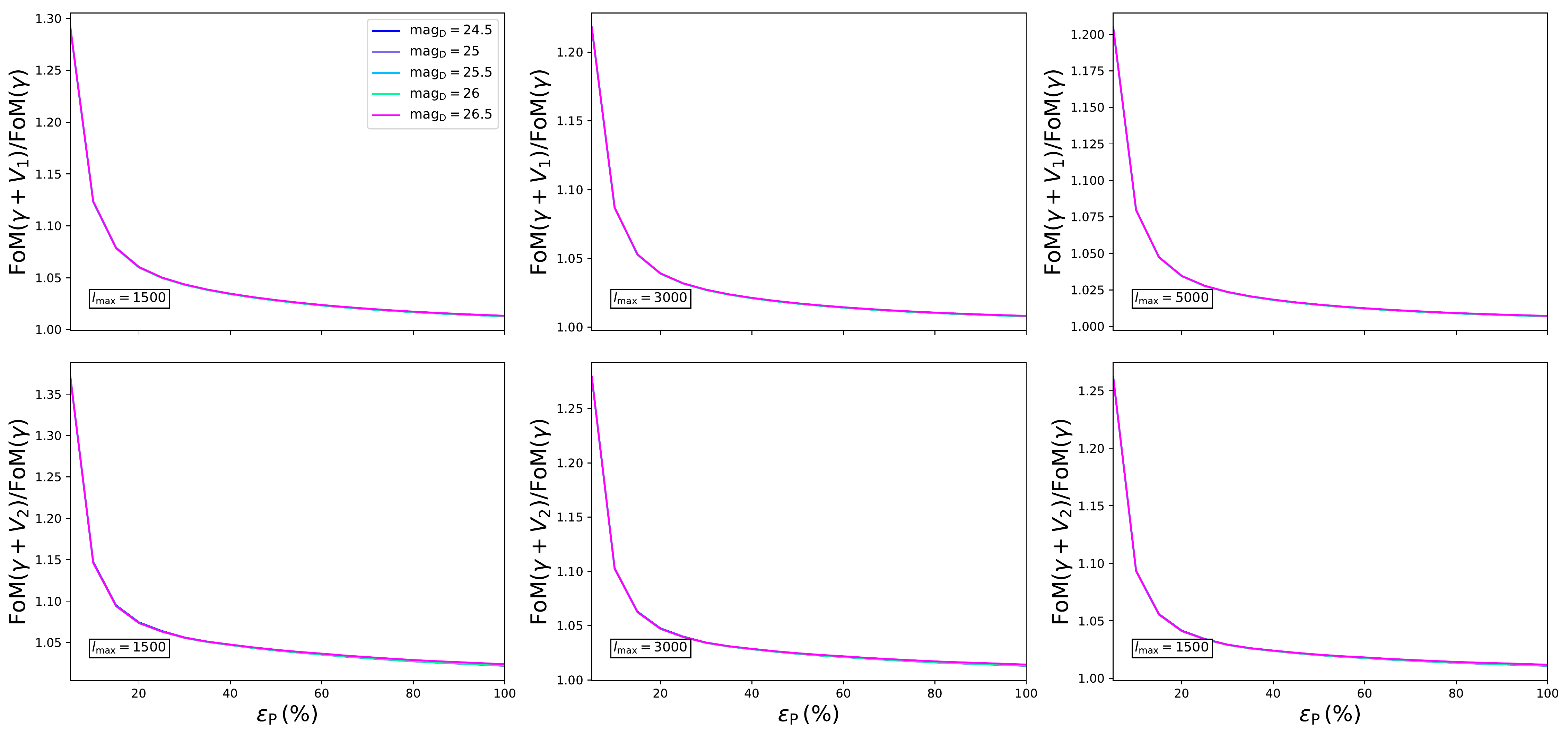}
\caption{\emph{Top}: FoM ratio as function of the prior $\varepsilon_{\rm{P}}$ on the MFs nuisance parameters for different values of the limiting magnitude $\rm{mag_D}$ of the deep survey, when adding $V_1$ only to the shear tomography with $\ell_{\rm{max}} = (1500, 3000, 5000)$ from left to right. \emph{Bottom}: same as in the top panel but for $V_2$ only added to the shear. Note that in each panel the curves for different ${\rm mag_D}$ are so superimposed that they cannot be seen at all.}
\label{fig: fomsingle}
\end{figure*}
\begin{equation}
{\bf F} = {\bf F}_{\rm{WL}} + {\bf F}_{\rm{MFW}} + {\bf F}_{\rm{MFD}} + {\bf P}
\label{eq: ftot}
\end{equation}
where ${\bf F}_{\rm{WL}}$ is the Fisher matrix for shear tomography on the full survey area, ${\bf F}_{\rm{MFW}}$ and ${\bf F}_{\rm{MFD}}$ are those for MFs from the wide and shallow and deep and narrow survey regions, and ${\bf P}$ is the priors matrix. We put priors on the nuisance parameters only so that ${\bf P}$ is a diagonal matrix with null values for the rows corresponding to cosmological parameters, and $(\varepsilon_{\rm{P}} \, p_{\rm{nuis}, \, i}^{\rm{fid}})^2$ for the rows referring to the nuisance ones. Varying $\varepsilon_{\rm{P}}$ will allow us to investigate to which accuracy the nuisance parameters should be known in order to improve the constraints on the cosmological parameters by a given factor. Actually, we will quantify this by looking at the FoM only since this is the quantity of interest to discriminate among rival dark energy models. 

A caveat is in order about Eq.(\ref{eq: ftot}). By summing up the Fisher matrices from the different probes, we are implicitly assuming that there is no correlation among the three probes. This is why we decided to evaluate separately the MFs on the wide and deep area so that they do not share any data hence being independent. On the contrary, the shear tomography is evaluated over the full survey area so that the same data are used for tomography and MFs. It is worth noting, however, that the two probes are radically different, with shear tomography probing the local properties of the shear field, and MFs the topological property of the full convergence map. Moreover, they are affected by different systematic effects and retrieved from different measurement pipelines so that one can argue that possible correlations (if any) are washed out by the estimate procedure. We therefore rely on Eq.(\ref{eq: ftot}) warning the reader that a definitive demonstration of its validity is a still pending issue.

We refer the reader to \citet{paper1} for the full set of formulae to compute the MFs Fisher matrix, but we stress here two remarkable differences concerning the inverse covariance matrix. This is still estimated as \citep{hartlap2007} 

\begin{equation}
{\bf Cov}^{-1} = \frac{N_{\rm f} - N_{\rm d} - 2}{N_{\rm f} - 1}\, ({\bf Cov}^{\rm obs} + {\bf Cov}^{\rm sys})^{-1}
\label{eq: covinv}
\end{equation} 
with the multiplicative factor that corrects for the finite number of realizations $N_{\rm f}$ used to estimate the covariance of the $N_{\rm d}$ dimensional data vector. In \citet{paper1}, we set $N_f = A/25$ with $A = \num{3500} \, \rm{deg}^2$ the total area cut from the MICEv2 simulated field. This limited the cases we could consider since one needs the multiplicative term to be positive. On the contrary, here, thanks to the use of \texttt{FLASK}, we have simulated a full sky survey that, after cuts to have well separated patches, provide us $N_{f} = \num{1108}$ subfields. This order of magnitude increase of $N_f$ and the smaller number of nuisance parameters (hence the smaller $N_d$), makes the multiplicative factor close to unity for all the cases of interest.

\begin{figure*}
\centering
\includegraphics[scale=.40]{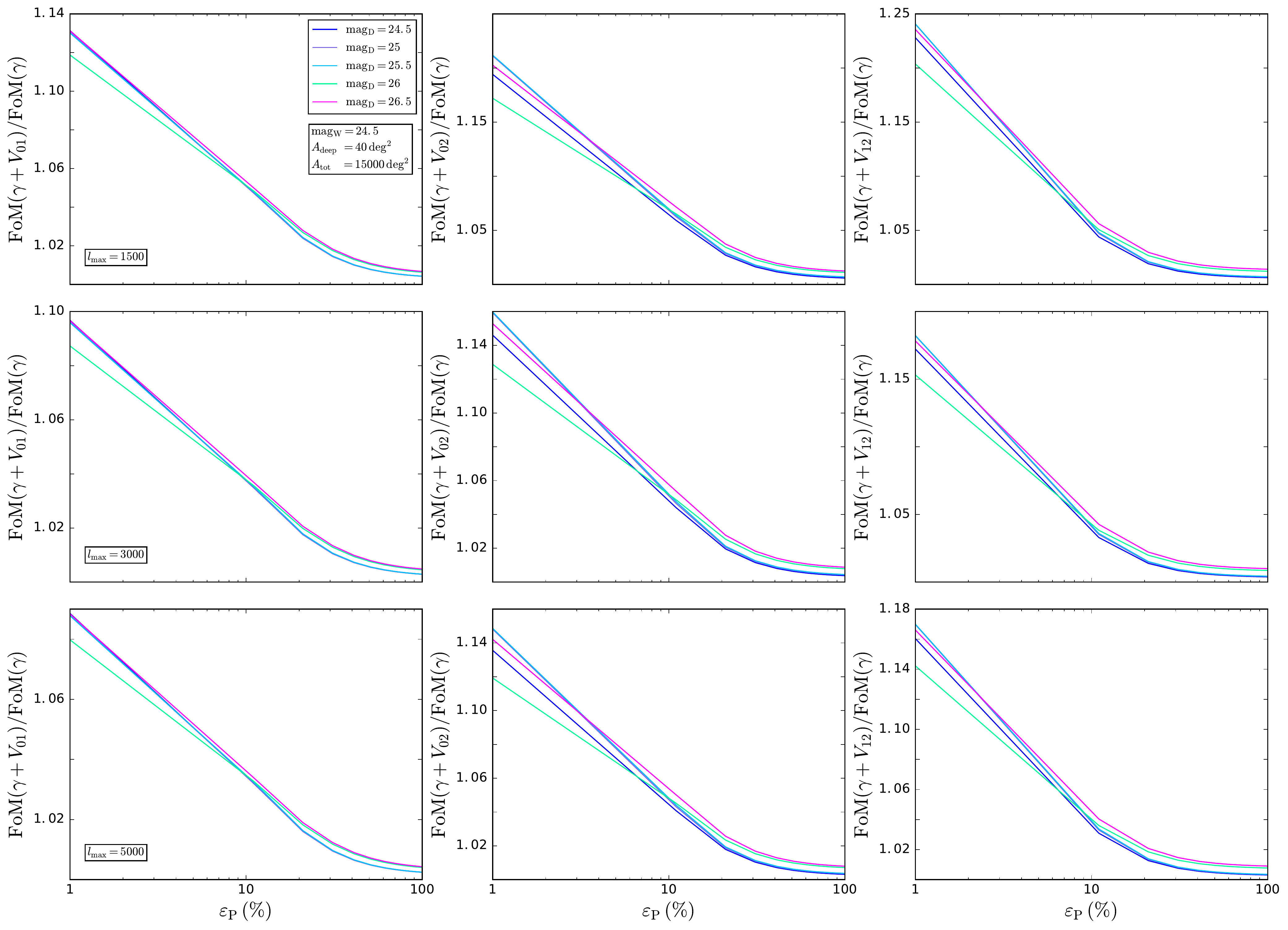}
\caption{Same as Fig.\,\ref{fig: fomsingle} but adding two MFs to shear tomography $V_{01}$, $V_{02}$, and $V_{12}$ for left, center, and right panels). Again, the dependence on ${\rm mag_D}$ is hard to appreciate so that, for most of the panels, it is impossible to see more than one curve. The only exception is the line referring to ${\rm mag_D} = 26.0$ in the last row bottom panels. Note that we plot here a smaller $\varepsilon_P$ range to better show the behaviour over the range where adding MFs to shear tomography indeed helps increasing the FoM by a significant amount.}
\label{fig: fomcouple}
\end{figure*}

Since the systematics covariance matrix is computed by propagating the errors on the nuisance parameters, setting a prior on ${\bf p}_{\rm{nuis}}$ will affect ${\bf Cov}^{\rm{sys}}$ too. In order to speed up the estimate, we first compute MCMC samples for each MFs dataset without any prior on ${\bf p}_{\rm{nuis}}$. When a prior is added, we perform importance sampling on the chains according to suitably defined Gaussian weights, thus recomputing the systematics covariance matrix entering Eq.(\ref{eq: covinv}). Note that the stronger is the prior, the smaller will be the contribution of ${\bf Cov}^{\rm{sys}}$. However, care must be taken to avoid the unrealistic case of ${\bf Cov}^{\rm{sys}}$ reducing to the null matrix. This can not be possible due to the approximated nature of our calibration formulae. We have checked, however, that as far as the prior is no smaller than $\sim 5\%$, ${\bf Cov}^{\rm{sys}}$ remains larger than ${\bf Cov}^{\rm{obs}}$ which is what we expect, given the large survey area we are considering.

\subsection{Figure of Merit improvement}

Adding MFs to shear tomography increases the number of observables, but also the number of nuisance parameters. Qualitatively, one can argue that the larger the number of probes, the stronger are the constraints. On the other hand, the larger the number of parameters, the weaker the constraints. Moreover, as shown in Fig.\,\ref{fig:mfs_cov}, there are strong correlations among MFs of different order at the same $(\theta_{\rm{s}}, \, z)$ so that it is worth wondering whether the use of a single MF is enough to improve the overall FoM. As a first test, we therefore investigate the ratio $\rm{FoM}(\gamma + V_n)/\rm{FoM}(\gamma)$ between the FoM from shear only and shear + MFs as a function of the prior on the MFs nuisance parameters. Hereafter, we will also consider three different shear only forecasts, which differ for the maximum multipole used in the forecasts. In particular, we set $\ell_{\rm{max}} = (\num{1500}, \, \num{3000}, \, \num{5000})$ for the pessimistic, intermediate, optimistic scenario.

We consider a $\num{15000} \, \rm{deg}^2$ survey with limiting magnitude $\rm{mag_{W}} = 24.5$, which includes a $40 \, \rm{deg}^2$ region observed at a deeper limiting magnitude $\rm{mag_{D}}$. As a general result, we find that adding $V_0$ only to the shear does not improve at all the FoM, no matter which prior is set on the nuisance parameters and which is the limiting magnitude of the deep field. The improvement is less than $10^{-4}$ so that we do not show the scaling of the FoM ratio with respect to $\varepsilon_P$. This is somewhat expected since $V_0$ only depends on the variance $\sigma_0$ of the convergence field. Since $\sigma_0$ is a second order quantity, it is not surprising that it does not add further information with respect to the one already probed by the more detailed second order statistics represented by the shear tomography. 

This is not the case for the higher order MFs $(V_1, \, V_2)$ that probe the non-Gaussianity of the convergence field. Since the number of nuisance parameters increases only by two when going from $V_1$ to $V_2$, it is expected that higher order probes have a larger impact on the FoM. This is indeed what the comparison of top and bottom panels in Fig.\,\ref{fig: fomsingle} shows. One could naively think that the FoM may be boosted a lot by the addition of a single MFs, either $V_1$ or $V_2$, given that the curves in the central and right panels go up to $\sim 20 - 30\%$. Unfortunately, such values are obtained only when $\varepsilon_P \sim 5\%$, while there is a steep decline in the range $(5, 15)\%$ followed by a shallow convergence towards a unit FoM ratio. Considering that, when no prior is used, the nuisance parameters are determined by the calibration procedure with roughly $60\%$ error, it is easy to understand that one should rather look at the part of the curve with $\varepsilon_P > 10 - 20\%$. In this regime, the FoM ratio is hardly improved by more than $\sim 5\%$. In particular, there is no appreciable dependence on the ${\rm mag_{D}}$ value as a likely consequence of the small contribution given by the ${\bf F}_{\rm{MFD}}$ to the sum in Eq.(\ref{eq: ftot}) when a single MF is used.

\begin{figure*}
\centering
\includegraphics[scale=.40]{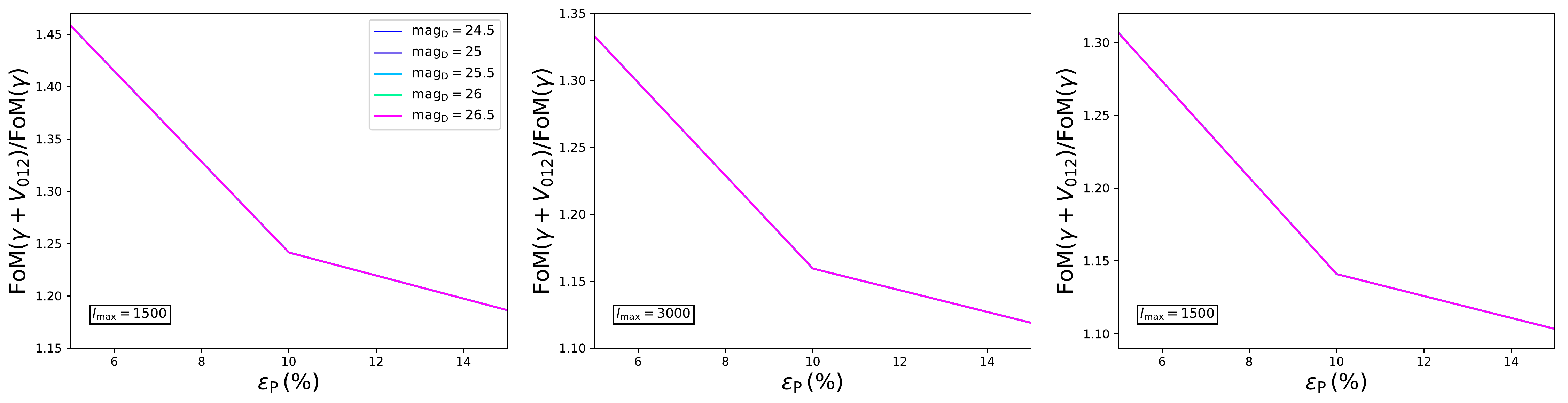}\caption{Same as Fig.\,\ref{fig: fomsingle} but adding all the three MFs to shear tomography with $\ell_{\rm{max}} = (\num{1500}, \num{3000}, \num{5000})$ from left to right. Curves for different ${\rm mag_{D}}$ values are again superimposed with the only difference of the green ones referring to ${\rm mag_D} = 24.5$.}
\label{fig: fomtriple}
\end{figure*}

We now move to Fig.\,\ref{fig: fomcouple} which shows the improvement in the FoM when two MFs are added to the shear tomography. We again find that the FoM ratio can get to surprisingly large values for $\varepsilon_P < 10\%$. However, significant improvement can still be obtained even for more realistic values of the priors with the FoM ratio being larger than $1.05$ for $\varepsilon_P$ values as large as $\sim 40\%$ for $V_{12}$. It is interesting to note that the trend of the FoM ratio with $\varepsilon_{\rm{P}}$ is roughly the same independently on which $\ell_{\rm{max}}$ value is used for the estimate of the FoM from shear tomography only, while it is only the scale of the $y$-axis in the different panels to change. This suggests that one can tailor the use of MFs as a way to partially compensate for a cut on $\ell_{\rm{max}}$. Such a shortening of the $\ell$ range can be of interest since the larger is $\ell$, the more one is pushing shear tomography into the highly uncertain nonlinear regime so that using a smaller $\ell_{\rm{max}}$ is a safer option to avoid theoretical errors due to inaccurate nonlinearities modeling. For instance, it is 
\begin{displaymath}
{\rm FoM}(\gamma, \, \ell_{\rm{max}} = \num{5000}) \simeq 1.19 \times {\rm FoM}(\gamma, \, \ell_{\rm{max}} = \num{3000}) \; ,
\end{displaymath}  
but we find that 
\begin{displaymath}
{\rm FoM}(\gamma, \, \ell_{\rm{max}} = \num{5000}) \simeq 
{\rm FoM}(\gamma + V_{12}, \, \ell_{\rm{max}} = \num{3000}, \varepsilon_P \simeq 8\%) \; ,
\end{displaymath}
while a $\sim 20\%$ prior is enough to half the FoM decrement due to going from $\ell_{max} = 5000$ to the safer $\ell_{max} = 3000$. Understanding which goal (i.e., better modeling nonlinearities vs improving MFs nuisance parameters priors) is easier to reach is a matter of open investigation. \\

\begin{figure*}
\centering
\includegraphics[scale=.40]{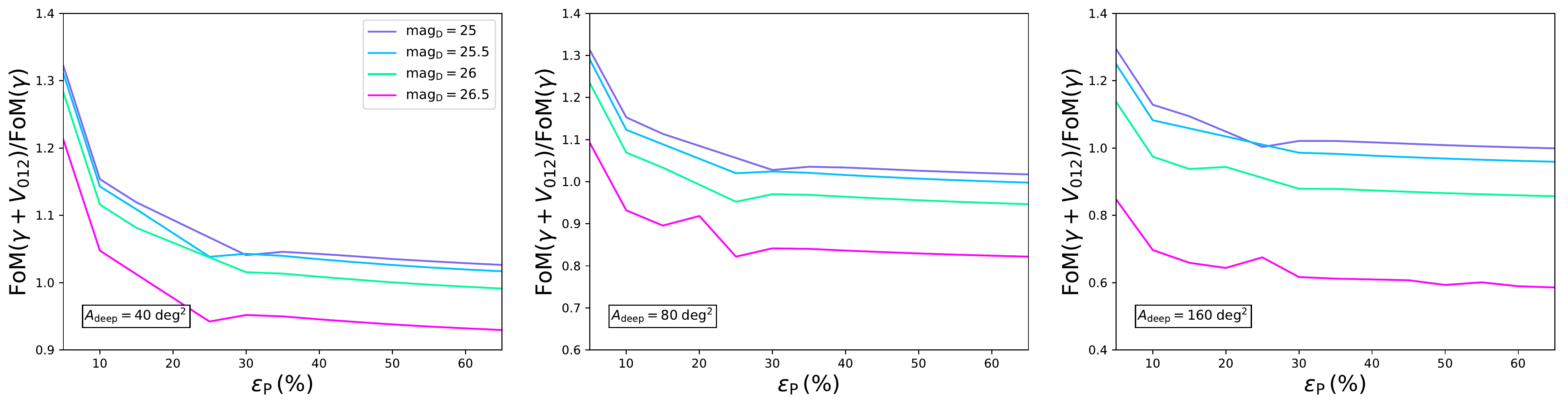}
\caption{\emph{Left}: FoM ratio as a function of the prior $\varepsilon_{\rm{P}}$ on the MFs nuisance parameters for the intermediate shear only scenario for different values of $(\rm{mag_{D}}, \, A_{\rm{tot}})$, using all three MFs. We fix $A_{\rm{deep}} = 40 \, \rm{deg}^2$. \emph{Center}: same as in the left panel but with $A_{\rm{deep}} = 80 \, \rm{deg}^2$. \emph{Right}: same as in the left panel but with $A_{\rm{deep}} = 160 \, \rm{deg}^2$. In each panel, blue, magenta, purple, and red lines refer to ${\rm mag_D} = (25.0, 25.5, 26.0, 26.5)$.}
\label{fig: fomvsmag}
\end{figure*}
Although the contribution of MFs is now more appreciable, we still find no significant dependence of the results on ${\rm mag_D}$. The curves in Fig.\,\ref{fig: fomcouple} are still superimposed so that they can not be discriminated. The only different case is the one corresponding to ${\rm mag_D} = 26.0$ in the $V_1 + V_2$ configuration. This could be related to some peculiarity in the calibration for this particular combination or an artifact of the importance sampling in the small $\varepsilon_P$ regime. We have been unable to undestand which hypothesis is the correct one, but we remark that the difference ony takes place in an unrealistic prior regime so that we do not care anymore. 

This is no more the case in Fig.\,\ref{fig: fomtriple} where we now use all three MFs for a joint analysis with shear tomography (for the three different $\ell_{\rm{max}}$ values). The only discrepant case is ${\rm mag_D} = 24.5$ (green curve) which actually refers to a configuration where there is no deep area at all since both the wide and the deep field have the same limiting magnitude.  We again find that MFs can compensate the FoM decrement due to the use of lower $\ell_{\rm{max}}$. We can also slightly relax the prior needed to get the same FoM as the optimistic shear only scenario since we find

\begin{displaymath}
{\rm FoM}(\gamma + V_{123}, \, \ell_{\rm{max}} = \num{3000}, \, \varepsilon_{\rm{P}} = 9\%) \simeq 
{\rm FoM}(\gamma, \, \ell_{\rm{max}} = \num{5000}) \; . 
\end{displaymath}
Still more interesting is to note that, for $\varepsilon_{\rm{P}} = 15\%$, we still get a combined FoM that is only $\simeq 6\%$ lower that the optimistic shear only one thus partially compensating the $19\%$ decrement observed due to cutting at $\ell_{\rm{max}} = \num{3000}$ rather than $\ell_{\rm{max}} = \num{5000}$. 

As a general result, we have found that there is almost no dependence of the FoM ratio on the limiting magnitude $\rm{mag_{D}}$ in the deep region. This is somewhat surprising since it could make us argue that there is actually no motivation for going deeper in magnitude. Actually, this is related to the approach undertaken in this paragraph where we have combined a wide survey with a deeper one which covers an area three orders of magnitude smaller.
The advantage of going deeper will become more eviden using the complementary approach explored in the next paragraph.
\begin{figure*}
\centering
\includegraphics[scale=.50]{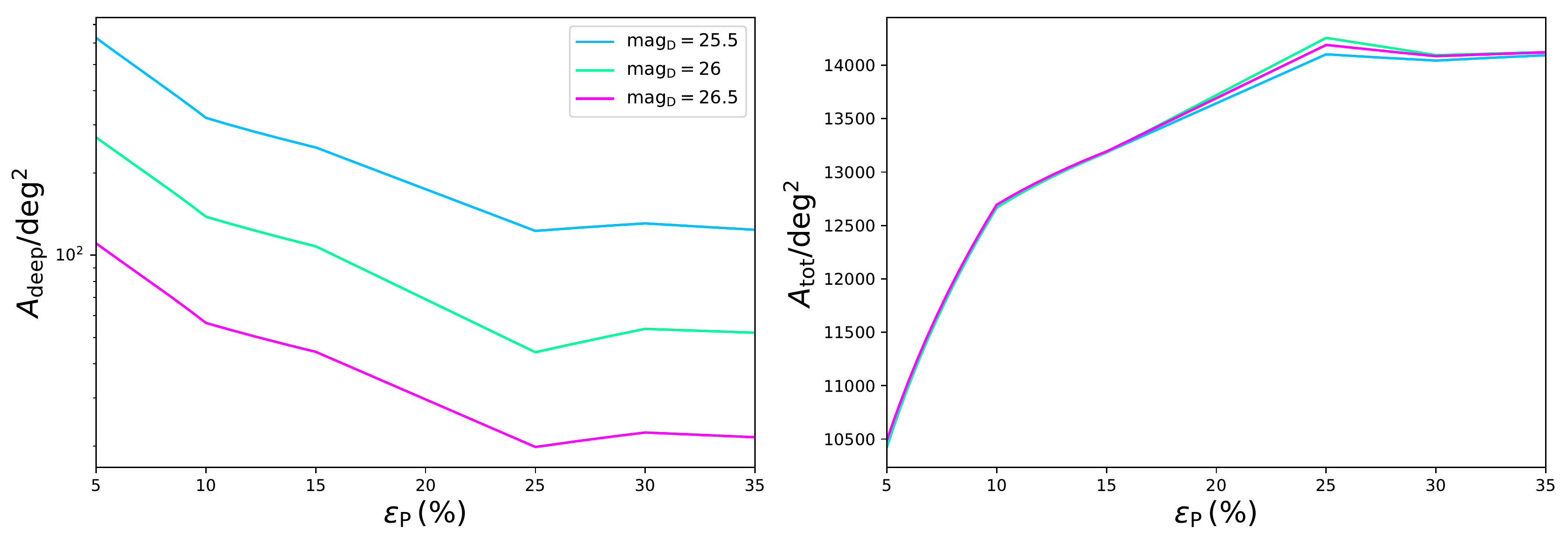}
\caption{\emph{Left}: deep survey area needed to get the same FoM as for the reference shear tomography only survey as a function of the prior on the MFs nuisance parameters for different limiting magnitude. \emph{Right}: same as in the left panel but considering the total survey area.}
\label{fig: deeptot}
\end{figure*}

\subsection{Optimizing an ideal survey}

In the previous subsection, we have considered the case of a survey with total area $A_{\rm{tot}} = \num{15000} \, \rm{deg}^2$, which includes a smaller portion $A_{\rm{deep}} = 40 \, \rm{deg}^2$ imaged at a deeper magnitude $\rm{mag_{D}}$. This choice is the same as the survey setup of the \Euclid mission. We here revert the point of view investigating how the shear + MFs FoM changes as a function of $(A_{\rm{tot}}, \, A_{\rm{deep}})$. We will hold fixed the total survey duration so that increasing $A_{\rm{deep}}$ is possible only at the cost of reducing the total survey area by a factor that depends on the chosen limiting magnitude $\rm{mag_{D}}$ (keeping fixed $\rm{mag_{W}} = 24.5$ for the wide area). It is important to stress that now we scale the results with respect to a reference FoM value, which is the shear tomography only one setting $\ell_{\rm{max}} = \num{3000}$ and $A_{\rm{tot}} = \num{15000} \, \rm{deg}^2$. We will only consider the case where all the three MFs are used since this is the one providing the largest increase in the FoM. 

This setup gives us the curves in Fig.\,\ref{fig: fomvsmag} where we show the FoM ratio as a function of the prior on MFs nuisance parameters for $\rm{mag_{D}}$ from 25.0 to 26.5 in steps of 0.5 and three different $A_{\rm{deep}}$  values. As a first remarkable result, we note that we can also have values of the FoM ratio smaller than unity, i.e., adding MFs reduces the overall FoM instead of increasing it. As counterintuitive as it can appear, this result is easily explained remembering that we are changing both $A_{\rm{deep}}$ and $A_{\rm{tot}}$. Since the shear only FoM linearly scales with $A_{\rm{tot}}$, going deeper over a large area can decrease the ${\bf F}_{\rm{WL}}$ term in Eq.(\ref{eq: ftot}) by an amount that is not compensated by the increase of the ${\bf F}_{\rm{MFD}}$ one. In such cases, the total FoM turns out to be smaller pointing in favor of a wide and shallow rather than deep and narrow survey. This is also confirmed by the fact that, for a fixed $\varepsilon_{\rm{P}}$, the FoM ratio typically\footnote{ Note that the curves in Fig.\,\ref{fig: fomvsmag} have a non smooth aspect since they have been obtained by interpolating over a grid in the $(\varepsilon, A_{deep})$ space. In order to save time, we have not used a grid fine enough to completely remove the numerical noise.} decreases with $\rm{mag_{D}}$ no matter which $A_{\rm{deep}}$ value is adopted. 

It is worth noting, however, that increasing the region imaged at a deeper magnitude can be desirable for other motivations indirectly related to the FoM validation (e.g., a better control of systematic errors thus making the forecasts more reliable) or to other aspects of the survey (such as the legacy outcome). It is therefore of interest to investigate whether it is possible to change $(A_{\rm{tot}}, \, A_{\rm{deep}})$ without affecting the total FoM. We therefore solve 
\begin{displaymath}
{\rm FoM}(\gamma + V_{123}, \, \varepsilon_{\rm{P}}, \, A_{\rm{deep}}) = {\rm FoM}_{\rm{ref}}
\end{displaymath}
with respect to $A_{\rm{deep}}$ for given $\varepsilon_{\rm{P}}$ fixing the total survey area in such a way that the survey duration is unchanged. The results are shown in Fig.\,\ref{fig: deeptot} for different values of the $\rm{mag_{D}}$. 

The curves in this plot may be used to optimize an ideal survey changing the areas of the deep and wide regions holding fixed the total duration. The answer depends on how well the MFs nuisance parameters are known. For instance, a $20\%$ prior on ${\bf p}_{\rm{nuis}}$ allows us to get the same reference FoM either for a survey with total area $A_{\rm{tot}} = \num{13466} \, \rm{deg}^2$ with $A_{\rm{deep}} = 437 \, \rm{deg}^2$ at $\rm{mag_{D}} = 25.0$, or by reducing $A_{\rm{deep}}$ to $36 \, \rm{deg}^2$ and $A_{\rm{tot}}$ to $\num{13544} \, \rm{deg}^2$ but with a significant deeper magnitude $\rm{mag_{D}} = 26.5$ (that may dramatically increase the legacy products).

Alternatively, Fig.\,\ref{fig: deeptot} may be used to put requirements on $\varepsilon_{\rm{P}}$ that should be fulfilled if one wants to increase the deep area at a given $\rm{mag_{D}}$. For instance, let us suppose we want to double the Euclid deep area, i.e., we set $A_{\rm{deep}} = 80 \, \rm{deg}^2$. In order to preserve the survey time duration, the total area should be reduced to 
\begin{displaymath}
A_{\rm{tot}} = \{\num{14719}, \, \num{14415}, \, \num{13652}, \, \num{11735}\} \ ,
\end{displaymath}
while the prior $\varepsilon_{\rm{P}}$  must be 
\begin{displaymath}
\varepsilon_{\rm{P}} = \{35, \, 31, \, 22, \, 7\}\% \ ,  
\end{displaymath} 
for ${\rm mag_{D}} = \{25.0, \, 25.5, \, 26.0, \, 26.5\}$ in order to preserve the same shear tomography + MFs FoM. Overall, Fig.\,\ref{fig: deeptot} tells us that, while it is indeed possible to reduce $A_{\rm{tot}}$ to enlarge $A_{\rm{deep}}$, the price to pay can be quite demanding. Indeed, $A_{\rm{tot}}$ quickly goes to its reference value as $\varepsilon_{\rm{P}}$ increases. A detailed analysis of the accuracy to which the MFs nuisance parameters may be constrained based on simulations is therefore mandatory, but outside the aim of the present paper.

\subsection{Uncertainty on the FoM ratio}

The results presented in the two previous paragraphs implicitly assume that the FoM is computed with no errors so that the ratio between the FoMs with or without the use of MFs can be reliably used to compare different setups. This assumption is motivated by the consideration that the FoM is estimated from the Fisher matrix which is a theoretical quantity so that, provided all the input ingredients are correct, it is known with infinite accuracy. However, one can wonder what if, for some unspecified reason, the Fisher matrix elements $F_{\alpha \beta}$ are incorrectly estimated. This can be the case because of either numerical errors or small discrepancies between the assumed survey setup and the final actual one\footnote{Ideally, $F_{\alpha \beta}$ can be radically different if one changes the assumed cosmological model or radically change the observational setup. However, these deviations should not be considered {\it uncertainties} to be propagated on $F_{\alpha \beta}$, but rather the Fisher matrix would refer to a different experiment and/or model so that it must deviate from the reference case.}. It is therefore worth wondering how such uncertainties propagate on the FoM ratio we have considered so far. 

To this end, let us assume that a given Fisher matrix element $F_{\alpha \alpha}$ has been estimated with an accuracy $\delta F_{\alpha \alpha}$. It has been shown \citep{IST2019} that the FoM is then known with an accuracy given by

\begin{equation}
\frac{\delta FoM}{FoM} = \frac{\delta F_{\alpha \beta}}{F_{\alpha \beta}} + 
\frac{F_{\alpha \alpha} F_{\beta \beta}/2}{F_{\alpha \alpha} F_{\beta \beta} - F_{\alpha \beta}^2}
\left ( \frac{\delta F_{\alpha \alpha}}{F_{\alpha \alpha}} + \frac{\delta F_{\beta \beta}}{F_{\beta \beta}}
- 2 \frac{\delta F_{\alpha \beta}}{F_{\alpha \beta}} \right ) 
\label{eq: deltafom}
\end{equation}
with $(\alpha, \beta)$ setting the column and row corresponding to the DE parameters $(w_0, w_a)$. In our case, the total Fisher matrix is the sum of three terms

\begin{figure}
\centering
\includegraphics[scale=.90]{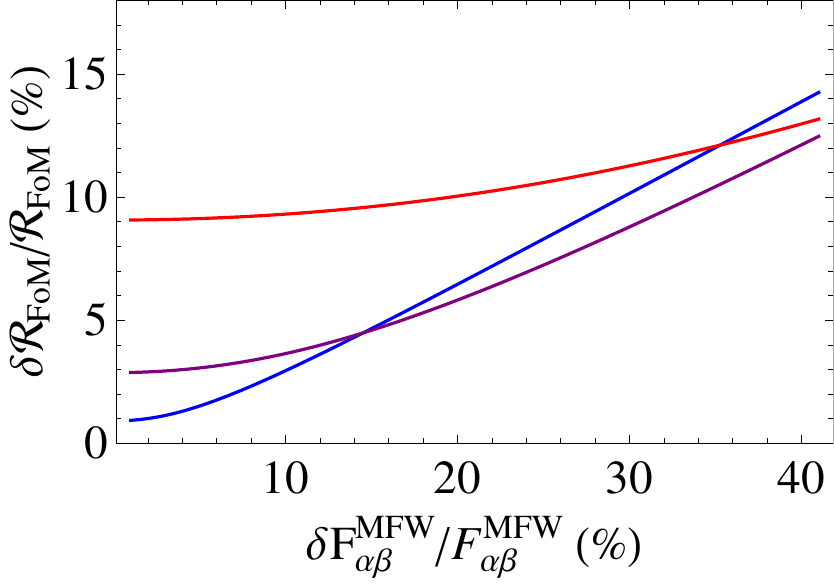}
 \caption{Error on the FoM ratio ${\cal{R}}$ as a function of the relative uncertainty on the MFW Fisher matrix elements. We set a $10\%$ prior on MF nuisance parameters, and consider the optimistic scenario for shear only. Blue, purple, and red lines refer to $\log{\varepsilon_{\alpha \alpha}^{WL}} = (-2.0, -1.5, -1.0)$ giving $\delta FoM(WL)/FoM(WL) = (0.6, 2.2, 6.4)\%$, respectively.}
\label{fig: errratio}
\end{figure}

\begin{displaymath}
F_{\alpha \beta} = F_{\alpha \beta}^{WL} + F_{\alpha \beta}^{MFW} + F_{\alpha \beta}^{MFD}
\end{displaymath}
which refer, respectively, to cosmic shear, MF on the wide area, MF on the deep area. Considering the three probes as independent, a naive propagation of errors gives

\begin{eqnarray}
\delta F_{\alpha \beta} & = & \left [ \left ( \delta F_{\alpha \beta}^{WL} \right )^2 
+ \left ( \delta F_{\alpha \beta}^{MFW} \right )^2  + \left ( \delta F_{\alpha \beta}^{MFD} \right )^2  \right ]^{1/2} \nonumber \\ 
 & = & \varepsilon_{\alpha \beta}^{WL} F_{\alpha \beta}^{WL} \nonumber \\
 &  \times &  \left [ 1 + 
\left ( \frac{\varepsilon_{\alpha \beta}^{MFW}}{\varepsilon_{\alpha \beta}^{WL}} \right )^2  
\left ( \frac{F_{\alpha \beta}^{MFW}}{F_{\alpha \beta}^{WL}} \right )^2  +
\left ( \frac{\varepsilon_{\alpha \beta}^{MFD}}{\varepsilon_{\alpha \beta}^{WL}} \right )^2  
\left ( \frac{F_{\alpha \beta}^{MFD}}{F_{\alpha \beta}^{WL}} \right )^2  \right ]^{1/2}
\label{eq: deltaftot}
\end{eqnarray}
with $\varepsilon_{\alpha \beta}^{X} = \delta F_{\alpha \beta}^{X}/F_{\alpha \beta}^{X}$. We can then use the other naive relation

\begin{equation}
F_{\alpha \beta} = F_{\alpha \beta}^{WL} \left ( 1 + \frac{F_{\alpha \beta}^{MFW}}{F_{\alpha \beta}^{WL}} +
\frac{F_{\alpha \beta}^{MFD}}{F_{\alpha \beta}^{WL}} \right )
\label{eq: fabvsfabwl}
\end{equation} 
to get an expression for $\varepsilon_{\alpha \beta} = \delta F_{\alpha \beta}/F_{\alpha \beta}$ and plug the result and Eq.(\ref{eq: fabvsfabwl}) into Eq.(\ref{eq: deltafom}) to ge the relative error on the FoM from the joint use of WL and MFs. Setting to zero the MFs terms gives the error on the FoM from WL only. We can finally write

\begin{equation}
 \frac{\delta {\cal{R}}_{FoM}}{{\cal{R}}_{FoM}} = \frac{\delta FoM(\gamma)}{FoM(\gamma)} 
\left \{ 1 + \left [ \frac{\delta FoM(\gamma + V_n)/FoM(\gamma + V_n)}{\delta FoM(\gamma)/FoM(\gamma)} 
\right ]^2 \right \}^{1/2}
\label{eq: deltafomratio}
\end{equation}
where we have defined ${\cal{R}}_{FoM} = FoM(\gamma + V_n)/FoM(\gamma)$ to denote the ratio among the FoM from WL+\,MFs and WL only, respectively. The relative errors $\delta FoM(X)/FoM(X)$ can be computed as described above and will lead to a lengthy yet simple algebraic formula (not reported here for sake of brevity) which provides the error on the FoM ratio as a function of the one on the WL only FoM and the relative uncertainties $(\varepsilon_{\alpha \alpha}, \varepsilon_{\alpha \beta}, \varepsilon_{\beta \beta})$ of the MFW and MFD Fisher matrices. 

In Fig.\,\ref{fig: errratio}, we plot $\delta {\cal{R}}_{FoM}/{\cal{R}}_{FoM}$ for the case of the optmistic cosmic shear scenario combined with MFs from wide and deep areas (for $\rm{mag_{lim}} = 25.5$ for the deep region) also adding a $10\%$ prior on MFs nuisance parameters. In order to reduce the number of parameters, we take the relative uncertainty on the WL Fisher matrix to be the same for $(\alpha, \beta)$ combination, and show the results as a function of $\delta F_{\alpha \beta}^{MFW}/F_{\alpha \beta}^{MFW}$ assuming the error on the other elements of both the MFW and MFD Fisher matrices are the same. Finally, for MFs, we use $(V_0, V_1, V_2)$ data. Note that dropping these assumptions does not qualitatively change the results with only a minor quantitative effect. 

This figure offers a qualitative way to set a requirement on the accuracy which the MFs Fisher matrix elements have to be determined with in order to trust the estimated value of ${\cal{R}}_{FoM}$. For instance, looking at the rightmost panel in Fig.\,\ref{fig: fomtriple}, we get ${\cal{R}}_{FoM} \simeq 1.15$ for the adopted prior on MFs. If we ask that ${\cal{R}}_{FoM} - \delta {\cal{R}}_{FoM} \simeq 1$ (i.e., we ask that the FoM improvement is larger than 1 at $1 \sigma$, we need to have $\delta {\cal{R}}_{FoM}/{\cal{R}}_{FoM} < 13\%$. Fig.\ref{fig: errratio} then tells us that this can be achieved as far as $\delta F_{\alpha \beta}^{MFW}/F_{\alpha \beta}^{MFW} < 35\%$. Although a detailed propagation of different errors on the input quantities has not been done, the margin is large enough to be confident that it can be fulfilled thus making our estimate of the FoM ratios quite reliable.

\section{Conclusions} \label{conclusions}

The higher sample size and the higher data quality promised by Stage IV lensing surveys make possible to go higher than second order statistics to probe the properties of the convergence field. Standard second order probes such as shear tomography power spectrum and two-points correlation function only trace the Gaussian properties of the field, while going to higher order allows us to probe its non-Gaussianity hence opening up the way to a better field description and hence stronger constraints on the underlying cosmological model. MFs stand out as promising candidates because of their dependence on the generalized skewness parameters probing the higher order statistical properties of both the field and its first derivative. In \citet{paper1}, we have shown how to match the theoretically predicted MFs based on a perturbed series expansion to the actually measured MFs on a convergence map reconstructed by noisy shear data. 

The present work differs from our previous paper in a number of aspects which makes a straightforward comparison not entirely possible. First, we have developed a novel calibration strategy that allows us to reduce the number of nuisance parameters. To this aim, we have derived, under reasonable assumption, the scaling of the noise-to-signal variance ratios and of the functions related to the skewness of the noise field. This derivation makes possible to halve the dimension of the nuisance parameters vector ${\bf p}_{\rm{nuis}}$ to 7 instead of the original 13. This significant decrease does not spoil down the quality of the matching between theory and data with the RMS of best fit residuals being almost the same as for the original recipe. In order to validate the scaling assumptions and determine fiducial values of the nuisance parameters, we have performed a joint fit to the full MF dataset thus taking into account the covariance among them. This is different from Paper I where we considered only the on diagonal elements of the covariance matrix. Such a more statistically correct approach also leads to us change the estimate of the systematics covariance matrix ${\bf Cov}^{{\rm sys}}$ which is now obtained by propagating the uncertainties in the determination of nuisance parameters on the final estimate of the theoretical MFs. As a  consequence, if a prior is added on the ${\bf p}_{nuis}$, ${\bf Cov}^{{\rm sys}}$ is accordingly changed reducing the impact on the overall error budget as expected. As a further improvement, we have also validated this calibration procedure against data with a different source redshift distribution and MFs S/N ratio, considering data sets at varying limiting magnitude $\rm{mag_{lim}}$.  

Two points remains to be still addressed. First, the validation has been carried out based on lognormal simulations generated with \texttt{FLASK} for a fixed set of cosmological parameters. Although the fiducial values used here are different from those in \citet{paper1}, it is critical to check that the proposed calibration procedure still holds in radically alternative cosmologies. By this, we do not mean that the nuisance parameters are the same, but that the set of Eqs.(\ref{eq: v0obs})--(\ref{eq: v2obs}) still allow us to match theory and data without dramatically increasing the RMS scatter of the best fit residuals since they enter the estimate of the total covariance matrix. \texttt{FLASK} is an ideal tool for such an analysis since it allows us to quickly generate convergence maps taking as input only the matter power spectrum for the given model. We are therefore planning to carry out a careful investigation of this issue also varying the number of maps and the noise properties. As a further step towards realistic mocks, we also plan to change the angular selection function in order to investigate the effect of the mask on the MFs measurement and the validity of the calibration procedure in this circumstance. Note that the impact of masking cannot be framed within the derivation of Eqs.(\ref{eq: v0obs})--(\ref{eq: v2obs}) so that it is possible that `ad hoc' corrections are needed. 

As a second step forward with respect to the first presentation of our approach to MFs in \citet{paper1}, we have here considered the more realistic case of a wide area survey imaged at a limiting magnitude $\rm{mag_{W}}$ containing a deep and narrow region with a larger limiting magnitude $\rm{mag_{D}}$. It turned out that a joint analysis of shear tomography and MFs (with contributions from both the wide and shallow and deep and narrow areas) may boost the total FoM. In particular, this allows us to reduce the maximum multipole $\ell_{\rm{max}}$ of shear tomography, partially compensating the loss in the shear only FoM thanks to the MFs contribution. Although we have carried on this analysis for a Euclid-like survey, we have also shown the requirements that should be set on the accuracy to which the MFs nuisance parameters have to be known in order to get the same FoM as the reference survey, but different values of the deep region area. Keeping unchanged the survey duration, an increase of $A_{\rm{deep}}$ comes at the cost of reducing $A_{\rm{tot}}$. MFs can then compensate the loss in FoM, opening the way to a different setup, which can help better controlling systematics and augmenting side products of great interest for the legacy science. 

As interesting as they can be, these results should nevertheless be taken `cum grano salis'. First, we have stressed that MFs complement and supplement shear tomography only if severe constraints on the nuisance parameters are available. It is a matter of open investigation to understand whether ${\bf p}_{\rm{nuis}}$ can indeed be constrained to the required accuracy. To this end, one should investigate how the error on the calibration procedure scales with the number of mock data sets. Moreover, one should also investigate whether the nuisance parameter accuracy scales with the noise properties, eventually setting requirements on this quantity too. We plan to address this issue in a forthcoming work relying on \texttt{FLASK} data under different cosmological scenarios to also check whether the full method works under all possible configurations (cosmology, noise, number of mock data sets, etc.). 

Another issue to be addressed concerns what is still missing in our framework. First, we have argued that the use of MFs can make possible to shorten the shear tomography multipole range thus being less dependent on an accurate modeling of the matter power spectrum in the highly nonlinear regime. However, this  implicitly assumes that MFs are less dependent on nonlinearities. Whether this is indeed the case is actually an open question, the hardest quantity to model being the matter bispectrum. However, this typically enters through a summation, which is weighted by the product of three exponential functions in $\ell$. High-$\ell$ terms are therefore strongly suppressed, making MFs likeley less sensible on the exact nonlinear recipe and to the impact of baryons. That this is indeed the case will be the subject of a forthcoming publication where we will compare whether the predicted MFs change when evaluated for the same cosmology but different approaches to model the effect of nonlinearities and baryons on the matter power spectrum and bispectrum.

A final ingredient missing is the intrinsic alignment (IA), which, in the weak regime, linearly adds to the lensing shear so that the reconstructed convergence field is a biased representation of the actual one. It is hard to qualitatively understand whether this has an impact or not on the estimate of MFs. On one hand, IA quickly becomes sub-dominant at high redshift so that a possible way out could be to cut the redshift range over which MFs are measured. Moreover, IA is a local effect that should not alter the global topology of the map hence again not affecting MFs. However, IA increases the correlation among close redshift bins hence possibly increasing also the correlation among MFs at different $z$, which decrease the MFs constraining power. Moreover, it is possible that IA works as an additional noise with its own properties (variance and generalized skewness) thus spoiling down the accuracy of the matching procedure between theory and data we have developed here. Although lensing simulations including the effect of IA are unavailable at the moment, one could investigate whether IA can be included in \texttt{FLASK} by using the option of generating the convergence field directly from a tomography spectrum including IA.

Summarizing, the present paper represents the second step along a path towards making MFs a common tool to be added to the standard second order shear statistics. As hard as the journey could be, we are confident that the final goal will be rewarding enough to compensate all the efforts to get there.

\section*{Acknowledgments}
CP and VFC are funded by Italian Space Agency (ASI) through contract
Euclid - IC (I/031/10/0)
and acknowledge financial contribution from the agreement
ASI/INAF/I/023/12/0. We acknowledge the support from the grant MIUR
PRIN 2015 Cosmology and Fundamental Physics: illuminating the Dark
Universe with Euclid.

\bibliographystyle{aa} 
\bibliography{mfdeep} 

\end{document}